\documentclass[aps,pre,twocolumn,floatfix,superscriptaddress]{revtex4-2}
\usepackage{amsmath} 
\usepackage{amsfonts}
\usepackage{braket}
\usepackage{graphicx}
\usepackage{wrapfig} 
\usepackage{gensymb}
\usepackage{color}
\usepackage{soul}
\usepackage{appendix}
\usepackage{bm}
\usepackage{csquotes}
\usepackage{enumitem}
\usepackage{physics}
\usepackage[svgnames]{xcolor}
\usepackage{hyperref}
\hypersetup{
  colorlinks=true,
  linkcolor=SteelBlue,
  urlcolor=SteelBlue,
  citecolor=ForestGreen
}

\graphicspath{ }

\newcommand{\uic}{Department of Mechanical and Industrial Engineering, University of Illinois Chicago, USA}
\newcommand{\ucphys}{Department of Physics, University of Chicago, USA}
\newcommand{\jfi}{James Franck Institute, University of Chicago, USA}
\newcommand{\leinweber}{Leinweber Institute for Theoretical Physics, University of Chicago, USA}

\newcommand{\lattvec}[1]{\mathbf{a}^{(#1)}}
\newcommand{\lattcomp}[2]{a^{(#1)}_{#2}}
\newcommand{\ulattvec}[1]{\hat{\mathbf{a}}^{(#1)}}
\newcommand{\ulattcomp}[2]{\hat{a}^{(#1)}_{#2}}

\newcommand{\lattvecperp}[1]{\mathbf{b}^{(#1)}}
\newcommand{\lattcompperp}[2]{b^{(#1)}_{#2}}
\newcommand{\ulattvecperp}[1]{\hat{\mathbf{b}}^{(#1)}}
\newcommand{\ulattcompperp}[2]{\hat{b}^{(#1)}_{#2}}

\newcommand{\bx}{\mathbf{x}}
\newcommand{\by}{\mathbf{y}}
\newcommand{\bq}{\mathbf{q}}
\newcommand{\br}{\mathbf{r}}
\newcommand{\bR}{\mathbf{R}}

\begin{document}
\title{Solids activated by multi-body nonreciprocity}
    
    \author{Brady Wu}
    \thanks{These authors contributed equally}
    \affiliation{\ucphys}
    \affiliation{\jfi}

    \author{Daniel S. Seara}
    \thanks{These authors contributed equally}
    \affiliation{\jfi}
    \affiliation{\uic}

    \author{Qinghao Mao}
    \affiliation{\ucphys}
    \affiliation{\jfi}

    \author{Vincenzo Vitelli}
    \affiliation{\ucphys}
    \affiliation{\jfi}
    \affiliation{\leinweber}

    \author{Heinrich M. Jaeger}
    \email{jaeger@uchicago.edu}
    \affiliation{\ucphys}
    \affiliation{\jfi}

    \date{\today}

\begin{abstract}
Field-mediated forces between particles can be both nonreciprocal and irreducibly multi-body, giving rise to emergent activity in systems of individually passive particles.
While prior work extensively studied how multi-body nonreciprocity can activate few-body systems, fuel intermittency, and drive liquid-like behavior, the effect on the solid state of large particle ensembles is less explored.
Here we use two-dimensional, close-packed rafts of identical, athermal, acoustically levitated microspheres to show how multi-body nonreciprocity injects activity in solids.
At the level of both individual particles and collective dynamics, crystalline rafts exhibit persistent circulations with no net handedness.
The rafts also exhibit a phonon-like dispersion relation corresponding to a large Poisson ratio, indicating an induced softness to shear.
Within the bulk, lattice symmetry masks microscopic nonreciprocity, although it can manifest in renormalized, reciprocal elastic coefficients.
At dislocations and raft edges, where this symmetry is broken, multi-body nonreciprocal interactions act as local generators of activity.
A coarse-grained microscopic model of multi-body nonreciprocity, calibrated on three-particle simulations, recovers the rafts' experimentally observed dispersion relations and localized activity.
Using these findings, we illustrate how to harness defect-induced activity for programmable movement by arranging microscopic dislocations to drive macroscopic propulsion and rotation of a whole raft.
\end{abstract}

\maketitle
\let\origaddcontentsline\addcontentsline
\renewcommand{\addcontentsline}[3]{}

\section*{Introduction}
An observation famously attributed to Sir Charles Frank is that \enquote{crystals are like people: it is their imperfections that make them interesting}~\cite{Berry2020}.
Here, we show that this holds even more true for solids with non-pairwise interactions that break reciprocity.
Reciprocity is a fundamental symmetry that arises across science, colloquially summarized as \enquote{$A$ affects $B$ as $B$ affects $A$}.
In many natural and engineered systems, reciprocity is broken and the interactions are said to be \textit{non}reciprocal~\cite{Fruchart2026}.
Nonreciprocal interactions can take many forms and lead to various consequences, from oscillations in models of magnetism~\cite{Avni2025} and flocking~\cite{Fruchart2021,Kreienkamp2022}, to traveling states in models of chemical reactions~\cite{Saha2020,You2020,FrohoffHuelsmann2023,Brauns2024}, catalytic colloids~\cite{OuazanReboul2023}, and human populations~\cite{Seara2025}.
In solids, nonreciprocal interactions give rise to novel material properties, such as odd elastic moduli~\cite{Scheibner2020}, breathing solitons~\cite{Veenstra2025a}, adaptive locomotion~\cite{Veenstra2025}, and motile dislocations~\cite{Braverman2021,Guillet2025}.

Several ingredients can create nonreciprocity.
Perhaps the most common route to nonreciprocity is to consider anisotropic forces, i.e. forces that depend not only on the distance between particles but also their orientation~\cite{Beatus2006,Ivlev2015,Yifat2018,Loos2022,Colen2024,Huang2024a}.
Another route is to consider interactions between non-identical particles~\cite{Martin2025,King2025,Morrell2026}.
Finally, forces can be non-central, i.e. the force need not point along the center-line between two particles~\cite{Bililign2021,Tan2022,Chao2026}, often induced by spinning particles~\cite{Soni2019,Reichhardt2026,Reichhardt2026a}.
Absent these ingredients---anisotropy, heterogeneity, or non-centrality---multi-body forces are necessary to break reciprocity.
This is seen in several systems, including robotic matter~\cite{Veenstra2025}, optical matter~\cite{Davenport2022,Parker2025,Linderman2026}, dusty plasmas~\cite{Thomas1994,Yu2025,Shu2026}, colloidal crystals~\cite{Zampetaki2021}, and acoustic levitation~\cite{St.Clair2023,Wu2023,Wu2025,Brown2026}.
Within the solid phase, multi-body nonreciprocity is known to drive intermittency~\cite{Shu2026} and melting~\cite{Wu2023}.
However, realizing stable solid phases and investigating the impact of multi-body nonreciprocity on the solid's properties, has received less attention outside of robotically programmed metamaterials with bond-bending interactions ~\cite{Scheibner2020,Veenstra2025}
and chiral biological tissues that offer little room for material control~\cite{Chen2025}. 
In solids composed of central, isotropic interactions, crystalline symmetry requires a different route for breaking reciprocity, namely using multi-body forces to exploit the spatial asymmetries generated by lattice imperfections. 
How such imperfections then can trigger both local and global activity in the solid is the focus of this work. 

Here, we systematically investigate the material properties of freely floating, two-dimensional (2D)  rafts composed of acoustically levitated microspheres with nonreciprocal multi-body interactions.
Within the levitation plane, particles interact via two sound-induced force fields: short-ranged repulsion due to hydrodynamic microstreaming and longer-ranged attraction due to acoustic scattering. 
These two interactions are conservative for $N=2$ particles, but once $N>2$, they contribute to nonreciprocity when configurational symmetry is broken~\cite{Lim2024}.
In the solid state of these rafts, where particles remain highly ordered and do not exchange neighbors, we find that multi-body nonreciprocal interactions nevertheless inject activity, evidenced by persistent circulations at the level of individual particles and collective modes.
To explain these observations, we (a) develop a minimal microscopic model of 3-body nonreciprocity, (b) calibrate it to experiments by extracting the relevant microscopic parameters, and (c) coarse-grain it to obtain elastic properties that are in agreement with macroscopic observations.
More specifically, the model recovers the dispersion relation measured from the power spectral density of lattice vibrations.
The model also predicts that the symmetries of our lattice hide the microscopic nonreciprocity, rendering the bulk a reciprocal, momentum-conserving solid.
However, there is an important caveat: microscopic nonreciprocity can renormalize the raft's elastic moduli, similar to a pre-stress.
It is at defects and other locations where the relevant spatial symmetries are broken (e.g. edges) that the effects of nonreciprocity are most significant.
Using computer simulations, we illustrate how to exploit this defect-induced nonreciprocal response to program motion in solids with multi-body forces beyond our model system.

\section*{Results}

\begin{figure*}[ht!]
    \centering
    \includegraphics{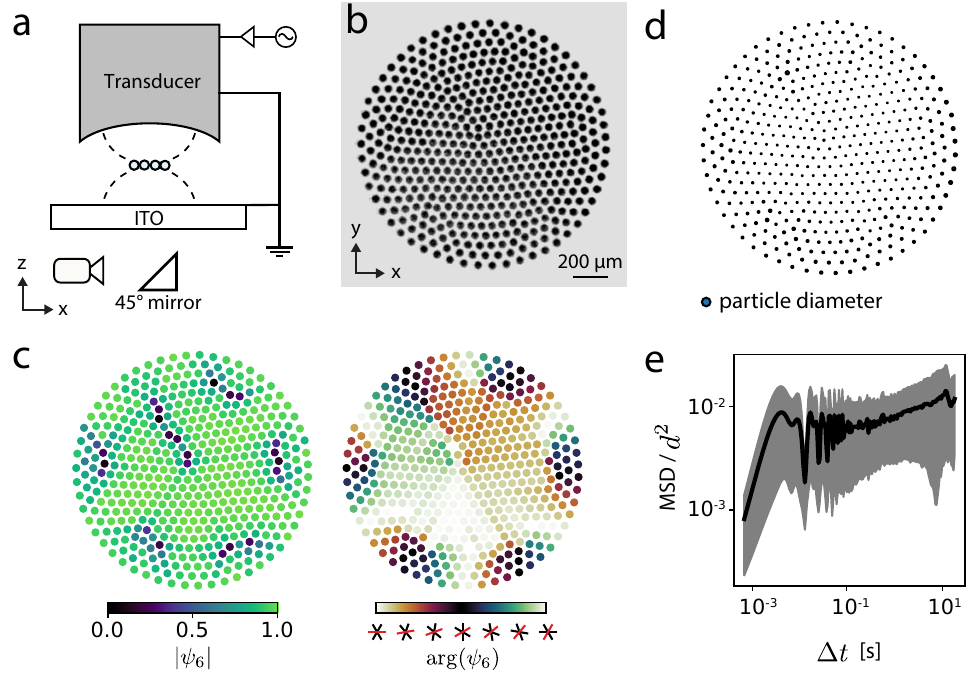}
    \caption{\textbf{Acoustic levitation creates a crystalline solid.}
    \textbf{a.} Sketch of the experimental setup.
    \textbf{b.} Bottom view of a raft of acoustically levitated particles.
    \textbf{c.} Magnitude (left) and argument (right) of the complex hexatic order parameter $\psi_6$.
    \textbf{d.} Square root of mean-square displacement (RMSD) of each particle, shown as a circle with radius given by the RMSD.
    The circle at the bottom indicates the particle size for the experiment in (b).
    \textbf{e.} MSD for the experiment in (b), showing a plateau at lag times $\sim 10^{-2}\,\mathrm{s}$.
    Line and shaded area show the mean $\pm$ standard deviation of the MSD over all particles.
    }
    \label{fig:expt}
\end{figure*}

\textbf{Creating a levitated crystal.}
Monodisperse microspheres of diameter $d = (41.1 \pm 0.5)\,\mu\mathrm{m}$ are acoustically levitated in air by ultrasound generated with a single-axis Langevin-type transducer. 
The transducer sets up a standing wave between its curved bottom surface and a reflecting glass plate, as in~\cite{Wu2023} (Fig.~\ref{fig:expt}a).
At drive frequency $f_d = 34.7\,\mathrm{kHz}$, the sound wavelength satisfies $\lambda \gg d$, and the experiments operate deep in the Rayleigh limit~\cite{Lim2024}.
The competition between short-range repulsion and longer-range attraction within the levitation plane produces monolayer rafts with a circular perimeter and a highly ordered interior (Fig.~\ref{fig:expt}b).
Dislocations comprising 5- and 7-fold defects mediate the transition from the triangular core to the circular periphery, reminiscent of a 2D Wigner crystal~\cite{Wigner1934,Radzvilavicius2011} (Fig.~\ref{fig:expt}c).
Unless otherwise noted, we focus on rafts with the particle size and drive frequency stated above.
We also include data for other values of $d$ and $f_d$.

Since both forces scale linearly with the sound energy density $E_0$, the particle separation does not depend on the sound intensity and instead is determined by the Stokes number $\Omega = \omega_d d^2/(4\nu_\mathrm{air})$, where $\omega_d = 2\pi f_d$ and $\nu_\mathrm{air}$ is the kinematic viscosity of the air surrounding the particles~\cite{Lim2024,Wu2023}.
As a result, individual microspheres maintain a well-defined average spacing from their neighbors that varies with position in the raft (Supplementary Figure 4).
We focus on small $E_0$, such that the raft is in an ordered, solid-like state without intermittent structural rearrangements~\cite{Brown2026,Wu2023}, as also seen with dusty plasma crystals~\cite{Shu2026}.
No particle moves more than one radius from its mean position (Fig.~\ref{fig:expt}d), and the mean-square displacement saturates within $\sim 10^{-2}$~s (Fig.~\ref{fig:expt}e).

\begin{figure*}[t]
    \centering
    \includegraphics{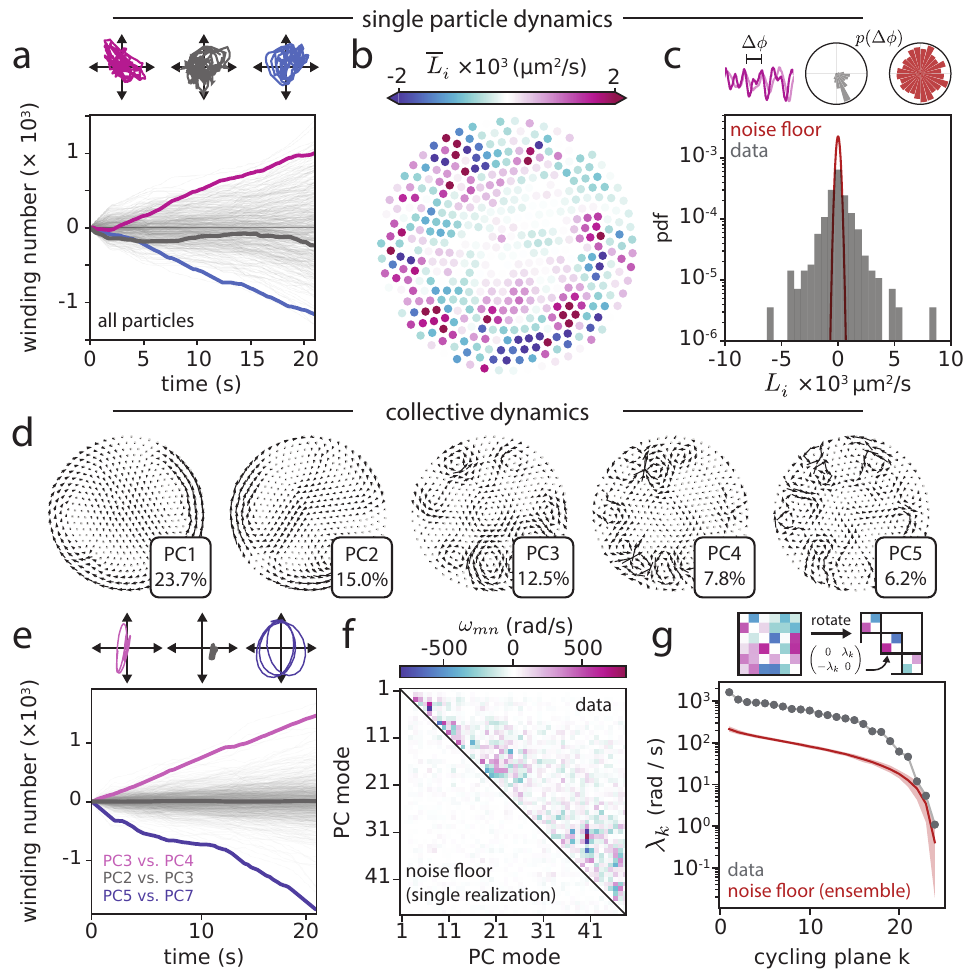}
    \caption{\textbf{Individual and collective activity in the rafts.}
    \textbf{a.} Cumulative winding number of all particles in the raft shown in Fig.~\ref{fig:expt}.
    Individual particles are highlighted that show large counter-clockwise rotations (pink), close to zero cumulative rotation (gray), and large clockwise rotations (blue).
    Trajectories for each of those particles over a period of 20 ms are shown above the plot.
    \textbf{b.} Spatial map of the time-averaged specific angular momentum ($\overline{L}_i$) of each particle, showing short-range correlation in particle rotation directions.
    \textbf{c.} Histogram of the specific angular momentum of the data (gray) and a noise floor (red).
    The noise floor is created by randomizing the phase between each particle's position and velocity ($\Delta \phi$), creating a surrogate time series with the same power spectrum but a uniform phase distribution, as illustrated above the plot.
    \textbf{d.} The five highest-variance PCA modes extracted from the particle position covariance matrix, labeled with the fraction of the total variance each carries.
    \textbf{e.} Cumulative winding number between all pairs of the first 48 PCA modes.
    Individual mode pairs are highlighted that show large counter-clockwise rotations (pink), near-zero cumulative rotation (gray), and large clockwise rotations (blue).
    \textbf{f.} Cycling frequency between every pair of the first 48 PCA modes, defined as $\omega_{mn} = \langle P_m \dot{P}_n - \dot{P}_m P_n \rangle / (2 \sigma_m \sigma_n) = -\omega_{nm}$, where $P_i(t)$ is PC$i$'s mode amplitude time series, and $\sigma_i$ is the standard deviation of $P_i(t)$.
    The lower triangle shows the result from a phase-randomized noise floor.
    \textbf{g.} Eigenvalue spectrum of the cycling matrix $\boldsymbol{\omega}$, showing a basis-independent measurement of the large rotations seen in the data compared to a phase-randomized noise floor.
    }
    \label{fig:phonons}
\end{figure*}

\textbf{Microscopic and collective nonreciprocity in the raft.}
Despite the small displacements of the particles, the raft is not in equilibrium.
Individual particles produce orbital trajectories around their mean position with persistent chirality, as measured using a cumulative winding number around each particle's mean position (Fig.~\ref{fig:phonons}a).
Across the raft, these orbits have no overall handedness, indicating that orbital chirality is spontaneously broken.
Such currents are generic signatures of nonreciprocal overdamped systems~\cite{Battle2016,Seara2018,Weiss2019}.
For underdamped systems, oscillations between conjugate variables are not forbidden, such as in the textbook example of position-velocity circulations in an underdamped harmonic oscillator.
To find activity in our underdamped system, we instead seek circulations that lie purely in position space.
Persistent circulations in position space are indeed forbidden in thermodynamic equilibrium (Supplementary Note V).
To quantify these circulations, we measure the specific angular momentum, i.e. the angular momentum normalized by mass, of each particle as $L_i = \mathbf{u}_i \times \dot{\mathbf{u}}_i$, where $\mathbf{u}_i = \br_i - \overline{\br}_i$ is the displacement of particle $i$ from its time-averaged position.
The temporal average of the specific angular momentum is correlated over short length-scales within the raft (Fig.~\ref{fig:phonons}b).
The magnitude of the observed angular momenta is significantly larger than a noise floor constructed by phase-randomized surrogate time series (Fig.~\ref{fig:phonons}c, Methods).
Taken together, these results show that the levitated raft is a solid whose individual constituents are driven away from equilibrium in a position-dependent manner.

This activity also arises in the collective motion of the raft.
We find collective modes by decomposing particle displacements using principal component analysis (PCA) (Fig.~\ref{fig:phonons}d, Methods).
Projecting the particle displacements into this space provides a time series of PC amplitudes, $P_i(t)$.
Pairs of PC amplitudes show persistent circulations, again measured using a cumulative winding number around $(P_m, P_n) = (0, 0)$ (Fig.~\ref{fig:phonons}e).
We compute the cycling frequency between two modes as
$\omega_{mn} = \langle P_m \dot{P}_n - \dot{P}_m P_n \rangle / (2 \sigma_m \sigma_n)$, where $\sigma_m^2 = \langle (P_m - \langle P_m \rangle)^2 \rangle$ is the variance of the time series $P_m(t)$. 
This defines a cycling-frequency matrix $\boldsymbol{\omega}$ with elements $\omega_{mn}$, which quantifies oscillations between pairs of the 48 highest-variance modes.
Pairs of PC modes exhibit large circulations across scales, well above a noise floor constructed from phase-randomized replicates of the PC-amplitude time series (Fig.~\ref{fig:phonons}f, Methods).
We further verify this using a basis-independent metric given by the eigenvalues of $\boldsymbol{\omega}$. These are purely imaginary and occur in complex-conjugate pairs because $\boldsymbol{\omega}$ is skew-symmetric, $\boldsymbol{\omega} = -\boldsymbol{\omega}^T$.
The magnitudes of the unique eigenvalues set a cycling-frequency scale in PC space and are an order of magnitude larger than the phase-randomized noise floor (Fig.~\ref{fig:phonons}g).

\textbf{A minimal model for multi-body nonreciprocity.}
In order to capture the nonequilibrium behavior of the raft discussed above, we construct the most general 3-body force consistent with the symmetries present in our system.
Our particles are identical, achiral, and passive.
Furthermore, the mediating medium is isotropic.
Thus, a 3-body force depends only on the shape of the triangle with the three interacting particles in its corners.
With $\hat{\br}_{ij}$ the unit vector from particle $i$ to particle $j$, $\hat{\br}^\perp_{ij}$ its counter-clockwise rotation, and $\delta r_{ij}=|\mathbf r_{ij}|-\ell$ its deviation from a reference spacing $\ell$, the force expanded to first order in bond deviations is
\begin{equation}
\label{eq:minimal_model}
\begin{split}
    \mathbf{F}_i = &\sum_{(j,k)} \left[a_{ijk} \hat{\br}_{ij} + b_{ijk} \hat{\br}^\perp_{ij}\right], \\
    a_{ijk} = &a_0 + k_a\delta r_{ij} + \\
    &a_3(\delta r_{ik}+\delta r_{jk}) + \\
    &\alpha_3(\delta r_{ik}-\delta r_{jk}),
\end{split}
\end{equation}
with $b_{ijk}$ the same form in the coefficients $b_0, k_b, b_3, \beta_3$.

\begin{figure*}[t!]
    \centering
    \includegraphics{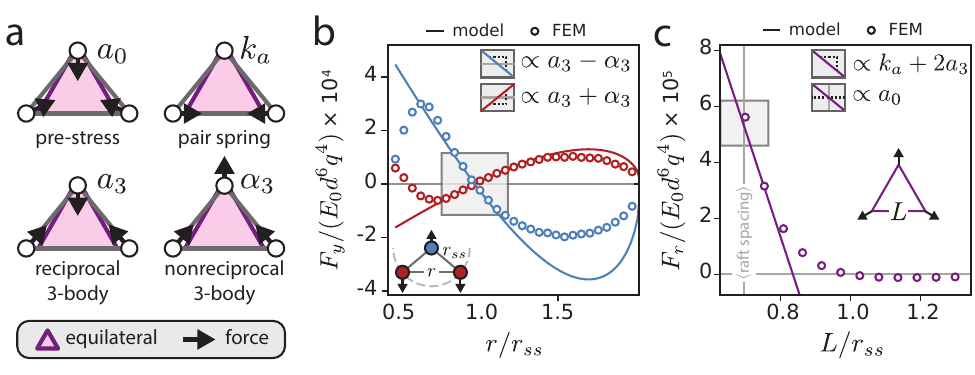}
    \caption{\textbf{Minimal model for nonreciprocal 3-body forces.}
    \textbf{a.} Illustration of the four terms found to be nonzero in the minimal model of Eq.~\eqref{eq:minimal_model}.
    \textbf{b.} The values of $a_3$ and $\alpha_3$ are measured from COMSOL simulations of three particles in an isosceles triangle configuration with changing opening angle of the triangle, parameterized by $r/r_{ss}$, where $r$ is the length of the base of the triangle and $r_{ss}$ is the side length of a force-free equilateral triangle and the length of the two equal-length sides of the isosceles triangle.
    The slope of the vertical component of the force on the particle at the apex of the triangle (blue) is given by $\partial_{r/r_{ss}} F_y^\mathrm{apex} |_{r/r_{ss}=1} = \sqrt{3} r_{ss} (\alpha_3 - a_3)$.
    The slope of the vertical component of the force on the two particles at the base of the triangle (red) is given by $\partial_{r/r_{ss}} F_y^\mathrm{base} |_{r/r_{ss}=1} = \sqrt{3} r_{ss} (\alpha_3 + a_3) / 2$.
    \textbf{c.} The values of $a_0$ and $k_a$ are measured from COMSOL simulations of three particles in an equilateral triangle configuration with varying side length $L$.
    The pre-stress is given by the value of the outward radial component of the force at the spacing observed experimentally, $F_r(\langle \ell\rangle) = -\sqrt{3} a_0$.
    The slope of the force at this point is given by $\partial_{L/r_{ss}} F_r|_{L = \langle \ell \rangle} = -\sqrt{3} r_{ss} (k_a + 2 a_3)$.
    }
    \label{fig:minimal_model}
\end{figure*}

Each of the coefficients has a clear interpretation: $a_0$ is the bond force at the reference spacing ($a_0 < 0$ corresponds to a compressive load), $k_a$ is a pairwise, Hookean stiffness, $a_3$ is a reciprocal 3-body force arising from the potential $U_3 = a_3(\delta r_{12} \delta r_{23} + \delta r_{12} \delta r_{13} + \delta r_{23} \delta r_{13})$, and $\alpha_3$ is a nonreciprocal, momentum-generating 3-body force (Fig.~\ref{fig:minimal_model}a).
The $b_{ijk}$ coefficients are the chiral counterparts to these forces.
For example, $k_b$ is a transverse spring stiffness that can give rise to odd elasticity~\cite{Scheibner2020}.
The net force on a particle triplet depends solely on $(\alpha_3, \beta_3)$ and a shape parameter that vanishes in an equilateral configuration (Supplementary Figure 2).

We measure these coefficients using forces extracted from finite element method (FEM) simulations of three levitated particles. The simulations resolve the full three-dimensional Navier--Stokes equations and acoustic scattering that generate the experimental forces.
We measure distances in units of the average interparticle separation, $\langle \ell \rangle \approx 54\,\mu\mathrm{m}$.
A sweep of isosceles triangles, with two sides held at $\ell$ and the third varied, isolates $a_3$ and $\alpha_3$ (Fig.~\ref{fig:minimal_model}b).
By contrast, sweeping over equilateral triangles yields $a_0$ and $k_a$ (Fig.~\ref{fig:minimal_model}c).
Reporting values as ratios to the Hookean spring constant $k_a$ (since the absolute force scale in the FEM simulations is not matched to experiment), we find
\begin{equation*}
\begin{split}
    \dfrac{a_0}{\ell k_a} &= -0.277 \pm 0.066, \\
    \dfrac{a_3}{k_a} &= 0.05 \pm 0.17, \\
    \dfrac{\alpha_3}{k_a} &= 0.0036 \pm 0.0008.
\end{split}
\end{equation*}
Note that the reciprocal 3-body force parameterized by $a_3$ is measured to be consistent with zero.
Particles in the lattice are more closely packed than an isolated three-particle system due to long-range attractive scattering forces, which introduce surface tension-like self-confinement that is extensive with particle number~\cite{Lim2022}.
Without explicitly modeling this self-confinement, it is absorbed into the pre-stress $a_0$.
Finally, we find the $b_{ijk}$ coefficients to be a factor of $\sim 10^{16}$ smaller than $k_a$.
The lack of chiral forces explains why the raft's circulations have no net handedness.
See Methods for details on fitting procedures.

\begin{figure*}[t]
    \centering
    \includegraphics{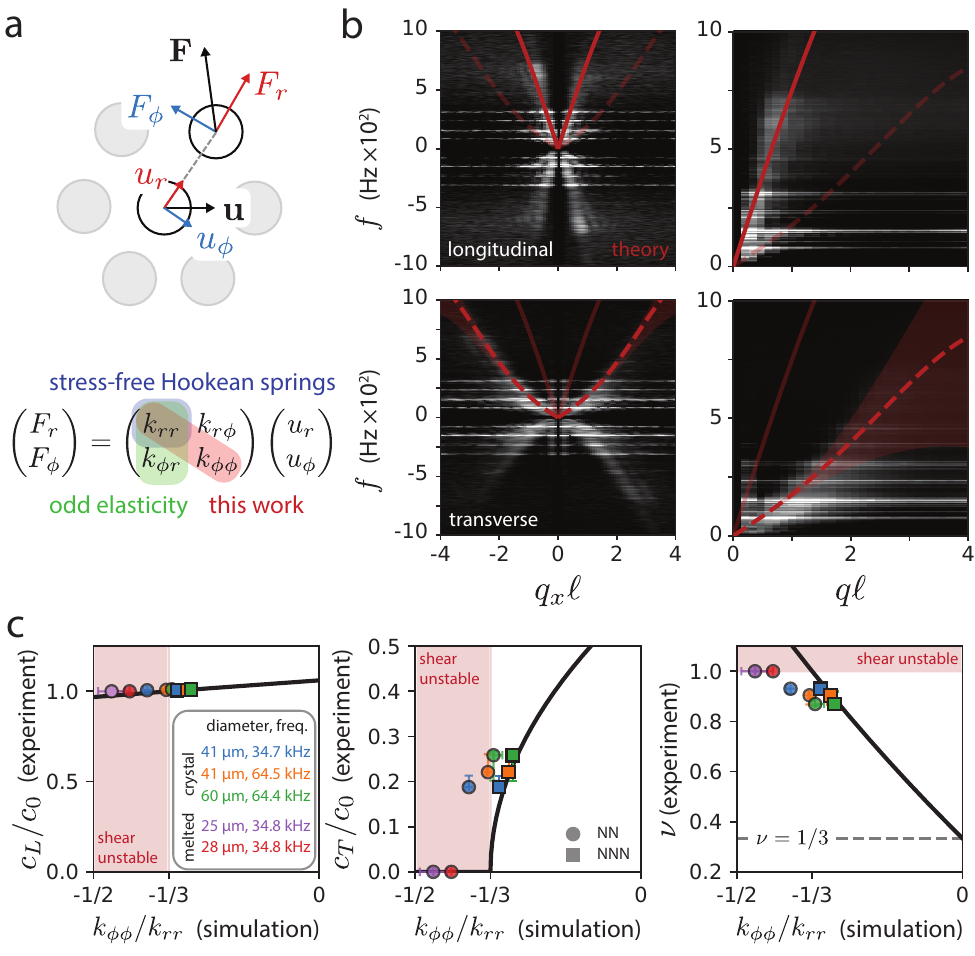}
    \caption{
    \textbf{Levitated rafts have anomalously large Poisson ratios.}
    \textbf{a.} The total force $\mathbf{F}$ induced by a small displacement $\mathbf{u}$ is decomposed into radial $F_r$ and azimuthal $F_\phi$ components.
    In this work, we find only $k_{rr}$ and $k_{\phi \phi}$ to be nonzero.
    \textbf{b.} Power spectral density in both longitudinal (top row) and transverse (bottom row) directions.
    The power spectrum is shown along $q_y = 0$ (left column) and as a function of the wavenumber $q = |\bq|$ (right column).
    Red lines show the longitudinal (solid) and transverse (dashed) dispersion relations of Eq.~\eqref{eq:linearForce} on a triangular lattice, fitted to the spectral ridges with two free parameters (Methods).
    Red shaded area for dispersion shows spread over $\pm10^\circ$ around $q_y=0$.
    \textbf{c.} Longitudinal speed of sound (left), transverse speed of sound (middle), and Poisson ratio (right) for experiments varying particle diameter and acoustic frequency.
    Speeds of sound are normalized by each raft's own fitted speed scale $c_0 = \ell\sqrt{k_{rr}/m}$, so that a single theory curve applies to all conditions.
    Circles show results from measuring $k_{\phi \phi}/k_{rr}$ using only nearest neighbors in FEM simulations, and squares show the result when considering next-nearest neighbors.
    Red shaded area shows the region of parameter space where $c_T^2 <0$, indicating a shear instability.
    }
    \label{fig:elasticity}
 \end{figure*}

\textbf{Multi-body nonreciprocity disappears in a perfect lattice.}
With a model in hand, we now turn toward understanding the bulk behavior of the raft.
Given the raft's crystalline structure, we derive the dispersion relation and elastic moduli due to our force law, Eq.~\eqref{eq:minimal_model}, when particles are placed on a triangular lattice.
Any force law linearized in a triangular lattice will result in a force parameterized by four coefficients, written in polar coordinates as
\begin{equation}
\label{eq:linearForce}
    F_i = - K_{ij} u_j; \quad
    K = \begin{pmatrix}
        k_{rr} & k_{r \phi} \\
        k_{\phi r} & k_{\phi \phi}
    \end{pmatrix},
\end{equation}
where $u_r$ and $u_\phi$ are the components of the relative displacement along and perpendicular to the bond axis, and $F_r$, $F_\phi$ the corresponding components of the force (Fig.~\ref{fig:elasticity}a).
A Hookean spring in a triangular lattice without compressive pre-stress would only produce $k_{rr} \neq 0$.
A non-central spring that generates both radial and transverse forces in response to radial strains has $\lbrace k_{rr}, k_{\phi r} \rbrace \neq 0$, which can give rise to odd elasticity~\cite{Scheibner2020}.

Our minimal model gives $\lbrace k_{rr}, k_{\phi \phi} \rbrace \neq 0$, written in terms of the minimal model coefficients as
\begin{equation}
\label{eq:Kmap}
\begin{split}
    k_{rr} &= 2 k_a + \dfrac{5}{2} a_3 - \dfrac{1}{2} \alpha_3, \\
    k_{\phi \phi} &= \dfrac{2 a_0}{\ell} - \dfrac{3}{2} a_3 + \dfrac{3}{2} \alpha_3.
\end{split}
\end{equation}
Note that this means $K$ is diagonal, indicating that nonreciprocity is invisible in the lattice, since activity would require off-diagonal elements of $K$.
Nevertheless, $\alpha_3$ still contributes to the force law, but only as a renormalization of the reciprocal linear coefficients $k_{rr}$ and $k_{\phi \phi}$.
Inserting the measured coefficients into Eq.~\eqref{eq:Kmap} gives $k_{\phi \phi}/k_{rr} = -0.29 \pm 0.09$.
A negative $k_{\phi \phi}$ means that displacing a neighbor transverse to a bond produces a transverse force that reinforces the displacement rather than opposing it.
This is equivalent to a nonrestorative torque, which we emphasize is a property of the force law evaluated in the lattice rather than a property of the minimal model.

\textbf{Renormalization of elastic moduli.}
We can then use this linear approximation to derive the speeds of sound and elastic moduli.
The elastodynamic equation for the displacement field $\mathbf{u}(\bx, t)$ is
\begin{equation}
\label{eq:elastodynamics}
    \rho \ddot{u}_i = F_i = \partial_j \sigma_{ij},
\end{equation}
where the stress $\sigma_{ij}$ is linearly related to the strain via the stiffness tensor, $\sigma_{ij} = C_{ijmn} \partial_m u_n$.
As Eq.~\eqref{eq:elastodynamics} is linear, it admits plane-wave solutions.
Expressing the equation in Fourier space in coordinates parallel ($u_r = \mathbf{u} \cdot \hat{\mathbf{q}}$) and perpendicular ($u_\phi = \mathbf{u} \times \hat{\mathbf{q}}$) to the waves, we get
\begin{equation}
\label{eq:elastodynamics_fourier}
    \rho \begin{pmatrix} \ddot{u}_r \\ \ddot{u}_\phi \end{pmatrix} =  -q^2 \begin{pmatrix} 
    B + G & 0 \\ 0 & G
    \end{pmatrix} \begin{pmatrix} u_r \\ u_\phi \end{pmatrix},
\end{equation}
where
\begin{equation}
    \label{eq:moduli}
    B = \dfrac{\sqrt{3}}{2}(k_{rr} - k_{\phi \phi}); \quad G = \dfrac{\sqrt{3}}{4} (k_{rr} + 3 k_{\phi \phi}).
\end{equation}
The azimuthal springs, which depend on the 3-body forces and the pre-stress, renormalize the bulk and shear moduli but do not introduce new moduli.
Importantly, the transverse forces drive the lattice toward liquid-like behavior with a shear modulus that decreases as $k_{\phi \phi}/k_{rr}<0$ increases in magnitude (Supplementary Note I.F).

We test this prediction against the measured vibrational spectrum of the raft, constructed as a power spectrum of velocity currents (Methods).
The spectrum is symmetric under $\bq \to -\bq$ (Fig.~\ref{fig:elasticity}b, left), precluding the parity symmetry breaking mechanisms that drive wave propagation in other active solids akin to a non-Hermitian skin effect~\cite{Beatus2006,Brandenbourger2019,Shankar2022}.
The spectrum shows pronounced peaks in both the longitudinal and transverse channels, which trace out the dispersion relation of the raft (Fig.~\ref{fig:elasticity}b, right).
We fit these peaks to the dispersion relation that Eq.~\eqref{eq:linearForce} produces on a triangular lattice (Supplementary Note II.A), with two free parameters per raft: the ratio $k_{\phi \phi}/k_{rr}$ and the speed scale $c_0 = \ell\sqrt{k_{rr}/m}$, where $m$ is the particle mass.
For the raft of Fig.~\ref{fig:expt} the fit returns $k_{\phi \phi}/k_{rr} = -0.302$, in good agreement with the earlier determination of Eq.~\eqref{eq:Kmap} based on an isolated trimer.

Another feature of the measured spectra is a set of horizontal lines superposed on the phonon branches (Fig.~\ref{fig:elasticity}b).
Each line is a single collective vibration that extends over a quarter to half of the particles (Supplementary Figure~5).
Its pattern varies irregularly from particle to particle, so its weight is spread across all wavevectors rather than concentrated at a single $q$.
The same particle-scale irregularity is visible in the PCA modes (Fig.~\ref{fig:phonons}d).
While we do not have a prediction for these lines, they are likely driven internally by the nonreciprocal interactions.
This is evidenced by the dynamical matrix in FEM simulations, which has its fastest-growing mode at the frequency of the dominant horizontal line (Supplementary Figure~5).

The $|\bq| \to 0$ limit of the dispersion relation gives the longitudinal and transverse speeds of sound.
In two-dimensional solids, they are related to the bulk ($B$) and shear ($G$) moduli by
\begin{equation}
\label{eq:speedOfSound}
    c_L = \sqrt{\dfrac{B+G}{\rho}}; \; c_T = \sqrt{\dfrac{G}{\rho}}.
\end{equation}
The ratio of the speeds of sound is related to the Poisson ratio $\nu$ via~\cite{Landau2009}
\begin{equation}
\label{eq:poisson}
    \nu = 1 - 2 \left(\dfrac{c_T}{c_L} \right)^2 = \dfrac{B-G}{B+G}.
\end{equation}

We repeat this procedure across five experimental conditions, varying the particle diameter from $25$ to $60\,\mu\mathrm{m}$ and the drive frequency from $34.7$ to $64.5\,\mathrm{kHz}$ (Fig.~\ref{fig:elasticity}c).
Three of the five conditions assemble into rafts that support phonons.
Their fitted $k_{\phi \phi}/k_{rr}$ ratios lie between $-0.274$ and $-0.302$, giving speeds of sound spanning $c_L = 0.16$--$0.34\,\mathrm{m/s}$ and $c_T = 0.035$--$0.088\,\mathrm{m/s}$, three to four orders of magnitude below the $343\,\mathrm{m/s}$ speed of sound in air at $20^\circ\mathrm{C}$ (see Supplementary Table 1 for exact values).

We then plot the measured speeds of sound against spring stiffness ratios $k_{\phi \phi}/k_{rr}$ obtained independently from FEM simulations of particles in a triangular lattice.
The data collapse onto the single curve predicted by Eq.~\eqref{eq:moduli} (Fig.~\ref{fig:elasticity}c).
Importantly, matching the data requires including next-nearest neighbors in the total force, highlighting the long-ranged nature of the multi-body nonreciprocal interactions captured by our minimal model.
The corresponding Poisson ratios are $\nu = 0.87$, $0.90$, and $0.93$, far above the $\nu = 1/3$ value for a typical triangular-lattice solid.

In the remaining two conditions, particles do not assemble into stable configurations, so there is no dispersion relation to fit. Since these conditions have no shear modulus, we assign them $c_T = 0$ and $\nu = 1$.

Taken together, the linear, two-dimensional elastic description of Eq.~\eqref{eq:linearForce} accounts for the anomalously high Poisson ratio of the rafts we observed.

\textbf{Multi-body nonreciprocity localizes to defects and edges.}
As the multi-body nonreciprocity is not apparent in a perfect lattice beyond renormalizing the elastic moduli, we next ask where activity in the raft originates.
Given the small displacements of the particles, we turn to the dynamical matrix $D_{ij}^{\alpha \beta}$ governing the raft's linear response (Methods).
Any dynamical matrix can be decomposed into a symmetric part and an antisymmetric part, $(D^\pm)_{ij}^{\alpha \beta} = (D_{ij}^{\alpha \beta} \pm D_{ji}^{\beta \alpha})/2$.
The dynamics derive from a potential if and only if $D^- = 0$.
We therefore take the per-particle ratio $\Vert D^-_i \Vert / \Vert D^+_i \Vert$ as a local measure of nonreciprocity, where $\Vert D^-_i \Vert^2 = \sum_{j \alpha \beta} [(D^-)_{ij}^{\alpha \beta}]^2$ is the Frobenius norm of the rows of $D^-$ corresponding to particle $i$, and similarly for $\Vert D^+_i \Vert^2$.

\begin{figure*}[t]
    \centering
    \includegraphics{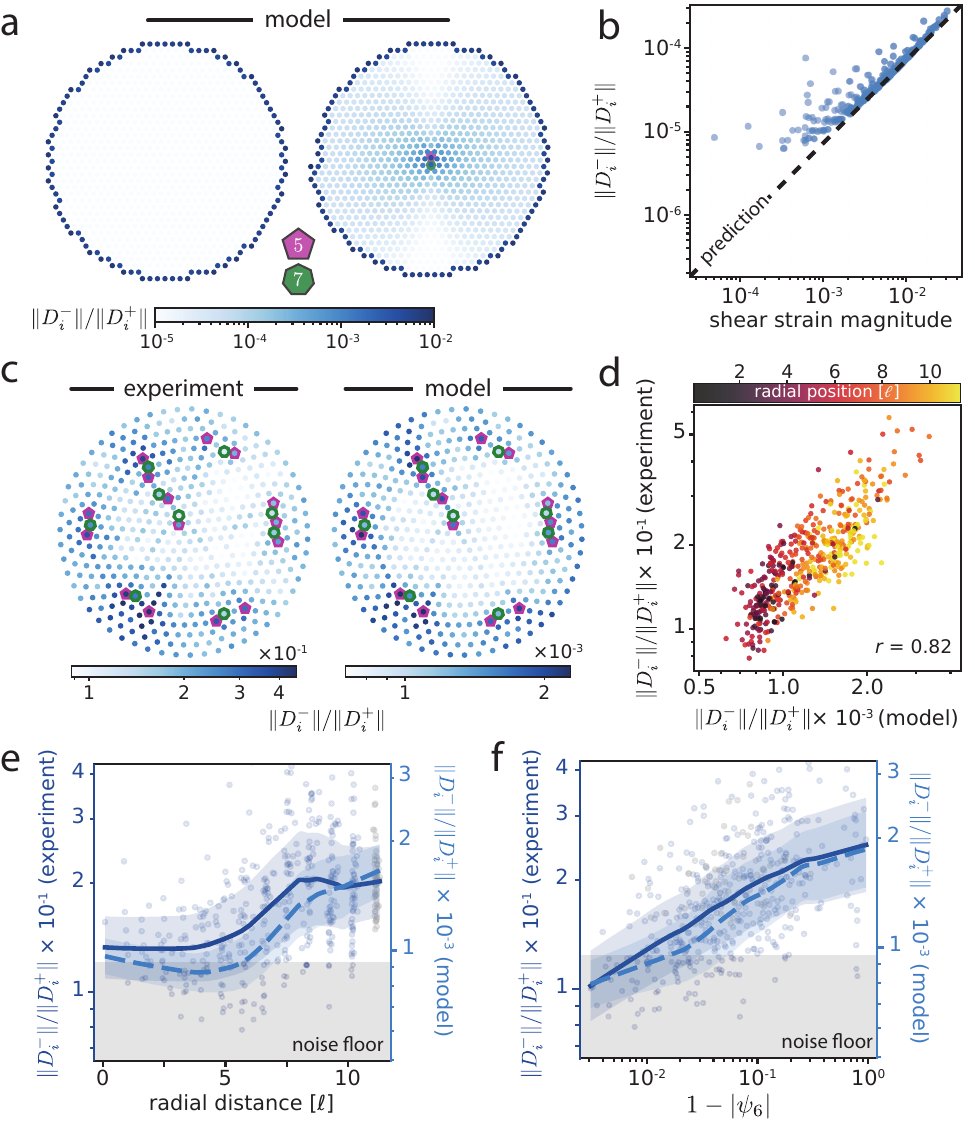}
    \caption{\textbf{Multi-body nonreciprocity emerges at defects and edges.}
    \textbf{a.} A raft of particles without (left) and with (right) a single $5$--$7$ dislocation pair in the center of the raft.
    Colors show the ratio $\Vert D^-_i \Vert / \Vert D^+_i \Vert$, a local measure of nonreciprocity, assuming the particles obey Eq.~\eqref{eq:minimal_model} with $\alpha_3 \neq 0$.
    Pink pentagons mark $5$-fold and green heptagons $7$-fold coordinated particles.
    Both disks share the color scale.
    \textbf{b.} The same ratio against the local shear strain magnitude for the dislocated raft.
    The dashed line is the analytic prediction, not a fit.
    \textbf{c.} Maps of the ratio for the experimental raft, from a dynamical matrix inferred from the measured trajectories (left, Methods), and for the minimal model evaluated at the same particle positions (right).
    Defects are marked as in (a).
    \textbf{d.} The two maps of (c) against each other, one point per particle, colored by distance from the raft centroid in units of the lattice constant $\ell$.
    \textbf{e.} The ratio against distance from the raft centroid, for the experiment (solid, left axis) and the model (dashed, right axis).
    Curves are local averages and shaded bands show $\pm$ the standard deviation.
    \textbf{f.} The same ratio against the hexatic order parameter, plotted as $1 - |\psi_6|$.
    The gray regions in (e, f) mark the noise floor of the experimental measure, constructed from reciprocal surrogates (Methods).
    Gray points are edge particles.
    }
    \label{fig:sources}
\end{figure*}

Simulating a raft with particles obeying Eq.~\eqref{eq:minimal_model}, we see that nonreciprocity arises when the six-fold symmetry of the triangular lattice is broken.
In a finite-sized raft with perfect lattice symmetry, $\Vert D^-_i \Vert / \Vert D^+_i \Vert > 0$ only at the edges (Fig.~\ref{fig:sources}a, left).
If a single edge dislocation is inserted into an otherwise perfect lattice, $\Vert D^-_i \Vert / \Vert D^+_i \Vert > 0$ throughout the raft (Fig.~\ref{fig:sources}a, right).
The dislocation produces a static, nonzero shear strain throughout the raft, and the local nonreciprocity grows proportionally to the shear strain magnitude, with a coefficient that we show analytically is generated by $\alpha_3$ alone (Fig.~\ref{fig:sources}b, Supplementary Note III.B).

We then turn to inferring $\Vert D^-_i \Vert / \Vert D^+_i \Vert$ in the experimental configurations.
We first construct an experimental dynamical matrix directly from trajectories using a time-lagged covariance matrix (Fig.~\ref{fig:sources}c, left; Methods).
We then compute the dynamical matrix in the experimental configuration from Eq.~\eqref{eq:minimal_model} using the coefficients inferred from the three-particle FEM simulations.
While the magnitudes of $\Vert D^-_i \Vert / \Vert D^+_i \Vert$ are different between the model and the experiment, their spatial organizations are highly correlated, giving a Pearson correlation coefficient of $r = 0.82$ (Fig.~\ref{fig:sources}d).

Looking more closely at the spatial organization of $\Vert D^-_i \Vert / \Vert D^+_i \Vert$, we see that it rises from the center of the raft to its edge (Fig.~\ref{fig:sources}e).
The experimental raft has fifteen 5-fold defects and nine 7-fold defects, mostly arranged near the edge of the raft.
Plotting $\Vert D^-_i \Vert / \Vert D^+_i \Vert$ as a function of $1 - \vert \psi_6 \vert$ shows that the activity indeed grows as the disorder increases, as quantified by a decreasing $| \psi_6 |$ (Fig.~\ref{fig:sources}f).

\textbf{Designing raft movement via lattice-gated activity.}
As a design rule, lattice-gated activity suggests a route to programmable motion in a raft.
This involves introducing dislocations at desired locations within an initially defect-free lattice.
While the current experiments do not allow for this, we provide two illustrative examples using simulations of crystalline rafts without external forces. Dislocations are seeded and the raft is then relaxed under central forces (Supplementary Note III.D).
Dislocations are placed with prescribed positions and orientations, quantified by their Burgers vectors $\mathbf{b}$.
When the particles obey Eq.~\eqref{eq:minimal_model} with $\alpha_3 \neq 0$, each dislocation experiences a climb force, perpendicular to $\mathbf{b}$, as follows from the mirror symmetry of the dislocation about its climb axis (Supplementary Note III.C).

\begin{figure*}[t]
    \centering
    \includegraphics{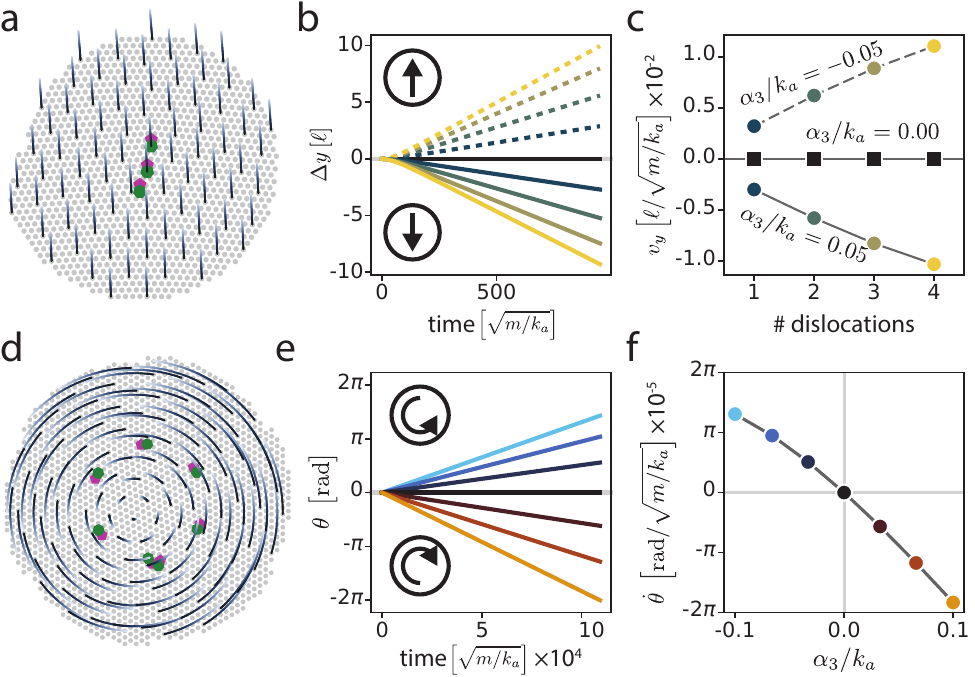}
    \caption{
    \textbf{Designing activity by dislocation placement.}
    Lattice units throughout: $\ell = 1$, $k_a = 1$ and particle mass $m = 1$.
    Every raft obeys underdamped dynamics, $m\ddot{\mathbf{r}}_i = \mathbf{F}_i - \gamma\dot{\mathbf{r}}_i$, with damping ratio $\zeta = \gamma/(2\sqrt{m k_a}) = 0.01$.
    Time is measured in units of $\sqrt{m/k_a}$ and displacement in units of $\ell$.
    Each raft has no external force, starts at rest from its relaxed $\alpha_3 = 0$ state, and $\alpha_3$ is switched on at $t = 0$.
    All other coefficients in Eq.~\ref{eq:minimal_model} are zero.
    \textbf{a}. A raft of radius $18\ell$ containing 3 dislocations with parallel Burgers vectors, stacked along the climb direction.
    5-fold and 7-fold defects are shown with magenta pentagons and green heptagons, respectively.
    Lines show the trajectories of a subset of particles while the raft travels $4\ell$ at $\alpha_3/k_a = 0.05$ (elapsed time $t \approx 530$).
    Time progresses from light to dark colors.
    \textbf{b}. Climb displacement $\Delta y$ against time for walls of one to four dislocations (dark blue to yellow), at $\alpha_3 / k_a = -0.05$ (dashed) and $+0.05$ (solid).
    The black line is $\alpha_3 = 0$.
    \textbf{c}. The resulting climb velocity $v_y$ against the number of dislocations in the wall, for $\alpha_3 / k_a = -0.05$, $0$ and $0.05$.
    \textbf{d}. A raft of radius $22\ell$ containing a ring of six dislocations at radius $r_0 = 9\ell$, whose Burgers vectors point radially outward with six-fold symmetry.
    Lines show the trajectories of a subset of particles over $30^\circ$ of rotation at $\alpha_3/k_a = 0.1$ (elapsed time $t \approx 9.2 \times 10^{3}$), progressing from light to dark.
    \textbf{e}. Orientation $\theta$ against time for the raft in (d), one curve per value of $\alpha_3 / k_a$ from $-0.1$ (light blue) to $0.1$ (orange).
    \textbf{f}. The angular velocity $\dot{\theta}$ against $\alpha_3 / k_a$.
    Colors correspond to the colors in panel (e).
    }
    \label{fig:engineered_activity}
\end{figure*}

The orientations of the Burgers vectors determine the macroscopic activity of the raft.
A tilt wall of $n$ dislocations with parallel Burgers vectors is mirror-symmetric about its climb axis.
That symmetry forbids both a torque and a glide force, leaving only a net force along the climb direction.
The seeded walls are only approximately mirror symmetric, and their residual glide grows from $1.4\%$ of the climb for one dislocation to $8.1\%$ for four (Supplementary Note III.C).
As a result, the raft propels at a speed that grows with $n$ and in a direction determined by the sign of $\alpha_3$ (Fig.~\ref{fig:engineered_activity}a--c).
By contrast, a ring of six dislocations whose Burgers vectors point radially outward has $\sum \mathbf{b} = 0$ and $C_6$ symmetry.
That symmetry forbids a net force but allows a net torque (Fig.~\ref{fig:engineered_activity}d).
As a consequence, the raft rotates (Fig.~\ref{fig:engineered_activity}e) in a direction determined again by the sign of $\alpha_3$ (Fig.~\ref{fig:engineered_activity}f).
In both designs, setting $\alpha_3 = 0$ leads to no motion.
While the simulations above have no pre-stress ($a_0 = 0$), the same designs under a uniform compression, equivalent to $a_0 \approx -0.08\,k_a \ell$, move in the same directions with rates changed by less than $20\%$ (Supplementary Figure~6, Supplementary Note III.F).

\section*{Conclusions}
Using acoustically levitated rafts of identical, athermal microparticles as a two-dimensional model solid, we combine experiments, simulations, and theory to show that multi-body nonreciprocal interactions have profound consequences.
In these underdamped solids, where particles interact through sound-field-mediated forces, nonreciprocity induces spontaneous particle motion and triggers propagating waves.
We introduce a minimal model that captures the small-amplitude vibrational dynamics by incorporating nonreciprocal three-body forces.
The model recovers the measured dispersion relations and shows how resistance to shear can be dramatically reduced, driving the Poisson ratio to a near-liquid value of $\approx 0.9$.

We find that activity is hidden in a perfect triangular lattice of identical achiral particles, in contrast to previously studied active solids comprising distinguishable constituents~\cite{King2025}, self-propelled nodes~\cite{Baconnier2022}, chiral interactions~\cite{Bililign2021}, or anisotropic interactions~\cite{Guillet2025}.
Instead, multi-body activity emerges at topological defects, where broken spatial symmetry drives local particle circulations and propagating waves. Defects therefore act as local generators of activity.
Finally, we demonstrate in simulations that this provides a design rule: macroscopic propulsion or rotation can be programmed into a raft by inserting microscopic dislocations that act as localized momentum sources.
Taken together, these results show how non-pairwise, nonreciprocal forces provide a tool set for activating solids and imbuing them with novel properties.

\section*{Methods}

\textbf{Acoustic levitation and imaging.}
Acoustic waves were generated using a standard Langevin transducer consisting of a piezoelectric element and an amplifying aluminum horn.
The bottom surface of the horn formed the upper boundary of the acoustic cavity in which the particles were trapped.
The horn had a concave bottom surface with a radius of curvature $R_c=25\,\mathrm{mm}$, which provided weak confinement of the particles toward the center of the acoustic trap.
The bottom surface was painted white to serve as a backlight for particle imaging when the cavity was illuminated from the side.
A flat reflector plate formed the lower boundary of the cavity and consisted of a glass substrate coated with transparent indium-tin-oxide, allowing imaging from below.
To ensure experimental repeatability, a heater tape wrapped around the horn and regulated by a temperature controller maintained the transducer at $(35.0\pm0.1)^\circ\mathrm{C}$.

The piezoelectric element was driven by a sinusoidal signal with a peak-to-peak voltage $V_{pp}$ ranging from 20 to 200 V, at a frequency close to the resonance frequency of the horn, $\omega_0/(2\pi)=34.870\,\mathrm{kHz}$.
The driving signal was generated using a function generator (EDU33212A, Keysight) and amplified by a high-voltage amplifier (A-301 HV amplifier, AA Lab Systems).
The acoustic pressure at the cavity perimeter was measured using a laser-based acoustic sensor (Xarion Eta100 Ultra optical microphone).
The gap between the horn and the reflector was adjusted close to $\lambda_0/2$, where $\lambda_0=9.8\,\mathrm{mm}$ is the wavelength corresponding to $\omega_0$.
This was achieved by vertically translating the horn using a translation stage.
Specifically, the driving frequency was fixed at $\omega_0$, and the gap height was adjusted until the measured acoustic pressure reached a maximum, indicating the resonance condition.
The gap height was kept constant throughout all experiments.

The particles consisted of polystyrene spheres (microparticles GmbH) with a material density of $\rho=1.05\,\mathrm{g/cm^3}$, a mean diameter of $d=41.11\,\mu\mathrm{m}$, and a diameter standard deviation of $0.5\,\mu\mathrm{m}$.
Particles were introduced into the cavity using a surgical probe with a tip diameter of approximately $100\,\mu\mathrm{m}$.
To minimize triboelectric charging, both the horn and reflector were grounded.
Any residual charge on the levitated particles was neutralized in situ at the beginning of each experiment by briefly applying soft X-rays using a Hamamatsu L12645 photoionizer.

To vary the particle size and drive frequency (Fig.~\ref{fig:elasticity}c), we additionally used polystyrene spheres from the same supplier with diameters $d=59.7\,\mu\mathrm{m}$ and $32.0\,\mu\mathrm{m}$, and a second horn resonant at $64.5\,\mathrm{kHz}$, with the gap tuned to $\lambda_0/2$ by the same procedure.
The particle diameters and drive frequencies of all conditions with measured spectra are listed in Supplementary Table~1.
All other procedures were identical to those described above.

All experiments were conducted in a temperature- and humidity-controlled laboratory environment, with a relative humidity of 45--50\% and an ambient temperature of 22--24$^\circ\mathrm{C}$.
The bottom reflector was cleaned with deionized water and dried with compressed air before and after each experiment.
Particle motion was recorded using a high-speed camera (Vision Research Phantom T1340) operating at 6,000 frames per second.
To minimize disturbances from ambient air currents, the entire experimental setup was enclosed within an acrylic box with dimensions of $61\times30\times46\,\mathrm{cm^3}$, substantially larger than the levitation region.

\textbf{Particle tracking.}
Particles were located in each frame of the high-speed recordings with a custom Python pipeline built on \texttt{OpenCV}, \texttt{scikit-image}, and \texttt{trackpy}~\cite{Crocker1996}.
A morphological black-hat filter with a $12\times12$~pixel square kernel removed the slowly varying background and enhanced the dark particle images.
The frame was then upsampled threefold by bicubic interpolation, and its intensity was rescaled to a fixed range before being thresholded into a binary image.
A morphological closing ($5\times5$ structuring element) filled small gaps within particle images.
Because neighboring particles in the raft were separated by gaps of only $\sim 0.3$ particle diameters, touching particle images were separated by a watershed segmentation of the Euclidean distance transform, seeded at its local maxima (minimum seed separation 8 upsampled pixels).
Each segmented region was assigned the position of its centroid and an equivalent radius $\sqrt{A_\mathrm{reg}/\pi}$, where the region area $A_\mathrm{reg}$ is unambiguous in the crystalline state.

We quantified the localization precision from the flat high-frequency floor of the displacement power spectrum, which corresponds to white positional noise of amplitude $0.18\,\mu\mathrm{m}$, less than $0.5\%$ of the lattice spacing.
Occasionally the segmentation resolved a single particle as two detections.
Detections closer than half the median nearest-neighbor distance were merged before structural analysis.
For the analysis of lattice vibrations, we restricted each recording to time windows in which the raft did not rearrange, identified as periods in which the Delaunay neighbor network of the time-averaged positions was unchanged.
Within these windows we removed the rigid-body motion of the raft: at each frame we subtracted the centroid of all particles, and we removed the global rotation by integrating the mean frame-to-frame change in the polar angle of all tracked particles about the centroid.
Displacements $\mathbf{u}_i(t)$ were then measured from each particle's time-averaged position in this co-rotating frame.

\textbf{Vibrational spectra and dispersion fitting.}
We extracted the dispersion spectra directly from the experiment using the procedure outlined in Ref.~\cite{Liu2010}.
We first computed the longitudinal and transverse currents, $j_x(q,t)=\sum v_{x,i}e^{-iqx_i(t)}$ and $j_y(q,t)=\sum v_{y,i}e^{-iqx_i(t)}$, where $x_i(t)$, $y_i(t)$ and $v_{x,i}(t)$, $v_{y,i}(t)$ are the $x$, $y$ position and velocity of particle $i$ at time $t$.
The phonon spectrum was computed by Fourier transforming the currents from $t$ to $\omega$, arriving at longitudinal $j_x(q,\omega)$ and transverse $j_y(q,\omega)$ power spectra.
Spectral ridges were located in both channels along the $\Gamma$--K and $\Gamma$--M directions, giving four branches per raft.
We fitted the nearest-neighbor lattice dispersion of Eq.~\eqref{eq:linearForce} to all four branches simultaneously with two free parameters, the ratio $k_{\phi \phi}/k_{rr}$ and the speed scale $c_0 = \ell\sqrt{k_{rr}/m}$, so that
$c_L/c_0 = \sqrt{(9 + 3k_{\phi \phi}/k_{rr})/8}$ and $c_T/c_0 = \sqrt{(3 + 9k_{\phi \phi}/k_{rr})/8}$.
Uncertainties on the ratio were obtained by bootstrap over ridge points.

\textbf{Finite-element simulations and coefficient fitting.}
We performed finite element method (FEM) simulations of particles levitating in an acoustic cavity using COMSOL Multiphysics\textsuperscript{\tiny\textregistered}.
The three-dimensional simulations accounted for both acoustic scattering and microstreaming effects in the surrounding air.
To obtain the time-averaged fluid properties, namely velocity and pressure, we first solved the linearized Navier--Stokes equations using the \texttt{Thermoviscous Acoustics, Frequency Domain} module, which accounts for viscous effects and resolves the acoustic boundary layer.
The frequency-domain formulation ensured that the resulting solution was free of transient effects.
The second-order time-averaged fluid properties were then computed by incorporating the appropriate source terms derived from the first-order solution into the \texttt{Laminar Flow} interface~\cite{Muller2014}.

The total force acting on each particle was calculated by summing the viscous and pressure contributions, which we refer to as the streaming and scattering components, respectively. Gravitational forces were neglected because gravity acts perpendicular to the levitation plane and therefore does not affect the in-plane motion of the particles. The particles and surrounding fluid were simulated within a rectangular domain of height $10d$ and width $40d$. The top and bottom walls were driven sinusoidally and in phase, with a velocity amplitude of $10^{-5}\,\mathrm{m/s}$ and a frequency of $34.7\,\mathrm{kHz}$. Slip boundary conditions were imposed on the lateral walls to minimize finite-size effects. The minimum and maximum mesh sizes in the vicinity of each particle were $0.6\,\mu\mathrm{m}$ and $9.5\,\mu\mathrm{m}$, respectively, providing sufficient resolution of the viscous and streaming effects. Finally, the COMSOL simulations were controlled through the \texttt{MATLAB} interface to perform parameter sweeps over different particle configurations.

Coefficients were obtained from two families of 3-particle configurations.
Sweeping isosceles triangles with the equal length sides held fixed separated $a_3$ from $\alpha_3$ through the slope of $F_y$ on the particles at the triangle apex and base (Fig.~\ref{fig:minimal_model}).
In contrast, sweeping equilateral triangles gave $a_0$ and $k_a$ through the intersection and slope of the radial force on all particles.
Each family was fitted at fixed leg length, since pooling across leg lengths is dominated by the steep decay of the force with separation.
Separately, we displaced a single particle of a perfect lattice in the finite-element model and recorded the resulting force on every other particle.
Projecting the first neighbor shell of that response onto the bond frame gave the bond-level ratio $k_{\phi \phi}/k_{rr}$, while inverting the nearest-neighbor lattice formula for the full multi-shell response gave the nearest-neighbor-equivalent ratio (Fig.~\ref{fig:elasticity}c).

\textbf{Dynamical matrices.}
Two independent routes were used, one for the model and one for the experiment.
For the model, we evaluated the definition
\begin{equation}
\label{eq:dynmat_def}
    D_{ij}^{\alpha \beta} = - \dfrac{\partial F_i^\alpha}{\partial x_j^\beta},
\end{equation}
by central differences on the force law of Eq.~\eqref{eq:minimal_model}, displacing one particle at a time by $h = 10^{-5} \ell$ and reading the resulting force on every particle.
The same construction was applied to every configuration we report: the defect-free raft, the raft containing a dislocation, and the measured particle positions.
Because it differentiates the implemented force law directly, this was used to independently check the analysis that produced Eq.~\eqref{eq:Kmap} (Supplementary Note I.F).
It also makes no assumption that the interactions derive from a potential, unlike approaches that estimate the dynamical matrix from a displacement covariance matrix~\cite{Kaya2010,Chen2013}, which cannot represent a nonreciprocal force law.

For the experiment, we worked in the basis of the softest normal modes and fitted the phase-space propagator
$\langle \mathbf{z}(t + \delta t)\mathbf{z}^\mathsf{T}\rangle / \langle \mathbf{z}\mathbf{z}^\mathsf{T}\rangle = e^{-\mathcal{J}\delta t}$,
where $\mathbf{z} = [\mathbf{u}, \dot{\mathbf{u}}]$ and
$\mathcal{J} = \left(\begin{smallmatrix} 0 & -I \\ D & \boldsymbol{\gamma} \end{smallmatrix}\right)$,
which returns $D$ in the mode basis without assuming reciprocity.
We verified that the inferred antisymmetric part is not produced by tracking noise (Supplementary Note IV).
The two or three softest modes lie below the acoustic dispersion and are dominated by the trap and the finite raft size.
As such, they were excluded from the analysis.

\textbf{Noise floors.}
Two surrogate constructions were used, both of which preserve the spectral content of the data while destroying the quantity being measured.
For the single-particle and collective circulations of Fig.~\ref{fig:phonons}, we randomized the phase between each particle's position and velocity, which leaves the power spectrum unchanged but makes the phase distribution uniform, and recomputed the angular momenta and cycling frequencies on the surrogate ensemble.
For the nonreciprocity maps of Fig.~\ref{fig:sources}, we built reciprocal surrogates that retain the per-mode spectral shape but carry no antisymmetric part.
The surrogates carry a uniform $2.4\%$ of each mode's power, an offset that does not enter the inferred $D$ (Supplementary Note IV).
The resulting band is a floor for the measured quantity only, and does not apply to the model.

\textbf{Engineered-activity simulations.}
The rafts of Fig.~\ref{fig:engineered_activity} were triangular-lattice disks with no external forces.
Dislocations were seeded by the continuum displacement field of an edge dislocation, and the raft was then relaxed under central springs of rest length $\ell$.
This relaxed lattice has Poisson ratio $1/3$.
The triangulation was then frozen, and the particles followed overdamped dynamics, $\gamma \dot{\mathbf{r}}_i = \mathbf{F}_i$, under the full force law of Eq.~\eqref{eq:minimal_model} at zero temperature.
Lengths are in units of the lattice constant, stiffnesses in units of $k_a$ and times in units of $\gamma/k_a$.
The net force and torque of the relaxed raft predicted its steady drift and rotation rates to within $5\%$.
In the walls, the Burgers vector lay to the side of the core, slightly breaking the mirror symmetry.
This effect compounded with additional dislocations, leading to a glide-to-climb ratio of $8.1\%$ for a wall of four.
In the ring, one of the six dislocations relaxed into a dislocation and an extra $5$--$7$ pair, which broke the six-fold symmetry and left a residual net force of $12\%$ of $|\tau|/r_0$, where $\tau$ is the net torque and $r_0 = 9\ell$ is the ring radius.
We additionally repeated the calculation with $\alpha_3$ allowed to vary with the mean side length of each parent triangle, since the three-particle simulations show that it depends on separation, and confirmed that the near-cancellation between triangles survives a non-uniform coefficient (Supplementary Note III.E).

\section*{Acknowledgments}
We thank Nina Brown, Michel Fruchart, Ella King, Billie Meadowcroft, Mia Morrell, and Bryan VanSaders for insightful discussions.
This research was supported by the Army Research Office through MURI grant W911NF-22-2-0109. Additional support was provided by a grant from the W.M. Keck Foundation.
The work utilized shared experimental facilities at the University of Chicago MRSEC, which was funded by the National Science Foundation under award number DMR-2011854.
The research also benefited from computational resources and services supported by the Research Computing Center at the University of Chicago.
This work was performed in part by D.S.S. at the Aspen Center for Physics, which is supported by National Science Foundation grant PHY-2210452.

\section*{AI Usage Declaration}
During the preparation of this work, the author(s) used Anthropic's Claude (Opus 5.5) in order to create simulation scripts, check mathematical calculations, and make figure scripts.
After using this tool/service, the author(s) reviewed and edited the content as needed and take(s) full responsibility for the content of the published article.

\bibliography{refs}

\newpage
\appendix
\onecolumngrid
\let\addcontentsline\origaddcontentsline
\begin{center}
    {\large \textbf{Supplement to: \\ Solids activated by multi-body nonreciprocity}}\\ [0.5em]
    { Brady Wu$^{1, 2}$, Daniel S. Seara$^{2, 3}$, Qinghao Mao$^{1, 2}$, Vincenzo Vitelli$^{1, 2, 4}$, Heinrich M. Jaeger$^{1, 2}$} \\ [0.5em]
    {\small \textit{
        $^1$\ucphys \\
        $^2$\jfi \\
        $^3$\uic \\
        $^4$\leinweber
    }}
\end{center}

\tableofcontents
\newpage

\section{Minimal model of the three-body force}
This section derives the minimal model introduced in the main text and its associated bond stiffness matrix and elastic moduli.
Throughout, $f$ and $g$ denote the scalar functions that set the longitudinal and transverse parts of the force; the coefficients $a_0, k_a, a_3, \alpha_3$ and $b_0, k_b, b_3, \beta_3$ are those of the main text.

Each particle has position $\br_i \in \mathbb{R}^2$.
We write the separation vectors as $\br_{ij} = \br_j - \br_i$, and the deviations from a reference length $\ell$ as $\delta r_{ij} = \vert \br_{ij} \vert - \ell = \delta r_{ji}$.
It is convenient to introduce a single-index shorthand $\delta r_i$ for the deviation of the triangle side opposite particle $i$:
$\delta r_1 \equiv \delta r_{23}$, $\delta r_2 \equiv \delta r_{31}$, and $\delta r_3 \equiv \delta r_{12}$.
With this labeling, $(\delta r_1, \delta r_2, \delta r_3)$ transforms as an $S_3$ permutation representation, i.e. relabeling the particles permutes the $\delta r_i$ in the same way.

\begin{figure}
    \centering
    \includegraphics{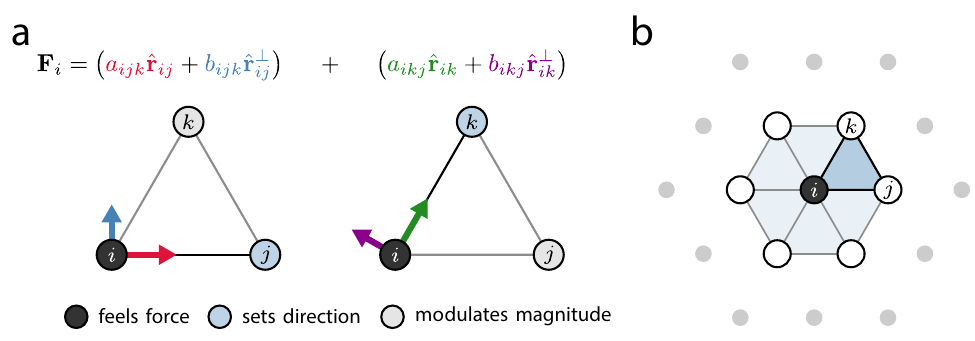}
    \caption{
        \textbf{Summation rule for the three-body force.}
        \textbf{a.} The two terms that a single triangle $\{i,j,k\}$ contributes to the force on particle $i$, Eq.~\eqref{eq:eom_general}.
        In the ordered pair $(j,k)$, particle $j$ sets the direction of the force (along $\hat{\br}_{ij}$ and its counter-clockwise rotation $\hat{\br}_{ij}^\perp$), while $k$ modulates its magnitude through the coefficients $a_{ijk}$ and $b_{ijk}$, which depend on all three side deviations $\delta r_{ij}$, $\delta r_{ik}$, $\delta r_{jk}$.
        The ordered pair $(k,j)$ swaps these roles.
        Arrow directions are schematic and do not imply signs of the coefficients.
        \textbf{b.} On the triangular lattice, site $i$ belongs to six nearest-neighbor triangles (shaded), each contributing two ordered pairs, for twelve terms in $\mathbf{F}_i$.
        The darker triangle is the one drawn in \textbf{a.}.
    }
    \label{fig:force_sum}
\end{figure}

We assume that the force is invariant under translations and permutations, and covariant under spatial rotations.
Under these assumptions, the most general three-body force on particle $i$ can be written as
\begin{equation}
    \label{eq:eom_general}
    \mathbf{F}_i = \sum_{( j, k )} a_{ijk} \hat{\br}_{ij} + b_{ijk} \hat{\br}_{ij}^\perp, \quad a_{ijk} = f(\delta r_{ij}; \delta r_{ik}, \delta r_{jk}), \; b_{ijk} = g(\delta r_{ij}; \delta r_{ik}, \delta r_{jk}),
\end{equation}
where the sum runs over ordered pairs $(j,k)$ and $(\hat{\br}_{ij}^\perp)_\alpha = R^{\pi/2}_{\alpha\beta}(\hat{\br}_{ij})_\beta$ is the unit bond vector rotated counter-clockwise by $\pi/2$.
The scalar functions $f(x;y,z)$ and $g(x;y,z)$ depend on three non-interchangeable arguments; accordingly, $a_{ijk}$ and $b_{ijk}$ are not symmetric under any exchange of two indices.
In this parameterization, $i$ labels the particle that feels the force, $j$ sets the force direction, and $k$ modulates the force magnitude (Fig.~\ref{fig:force_sum}a).

\subsection{Conservation laws}
We next ask: what general features of Eq.~\eqref{eq:eom_general} lead to dynamics that conserve energy, linear momentum, and angular momentum?

\textbf{Energy.} The force law is conservative if $\mathbf{F}_i = - \nabla_{\br_i} U$ for some potential $U$.
Any such $U$ must inherit the symmetries of the force law, and for identical particles this already forces $U = U(\delta r_{12}, \delta r_{13}, \delta r_{23})$.
To see this, note that the most general translation- and rotation-invariant function of three points in the plane is $U = U_+(\delta r) + \sigma \, U_-(\delta r)$, where $\sigma = \mathrm{sign}(\mathcal{A})$ is the orientation of the triangle; the magnitude $|\mathcal{A}|$ is fixed by the three side lengths through Heron's formula, so it adds nothing new.
Permutation invariance then requires $U_-$ to be totally antisymmetric in $(\delta r_1, \delta r_2, \delta r_3)$, since $\sigma$ changes sign under odd permutations.
Finally, $\sigma$ is locally constant away from collinear configurations, so $\nabla_{\br_i} (\sigma U_-) = \sigma \nabla_{\br_i} U_-$, which points along bond directions like every other term.
Every conservative force in this framework is therefore central, and in particular
\begin{equation}
    \label{eq:conservative}
    a_{ijk} = \dfrac{\partial U}{\partial \delta r_{ij}}; \quad b_{ijk} = 0 .
\end{equation}
Note that the right-hand side does not depend on $k$ and is symmetric under exchange of $i \leftrightarrow j$.
Let us define $a_{ijk}^\pm = (a_{ijk} \pm a_{jik})/2$ as the $(ij)$-symmetric and antisymmetric portions of $a_{ijk}$, which amounts to the symmetry arising from swapping the last two arguments of $f$ (equivalently for $g$).
The first expression in Eq.~\eqref{eq:conservative} then requires $a_{ijk}^- = 0$.

However, $a_{ijk} = a_{ijk}^+$ and $b_{ijk} = 0$ are necessary but not sufficient.
They are consequences of assuming that $U$ exists; nothing so far guarantees that it does.
What remains is an integrability condition, and it is easiest to see in shape space.
For a reciprocal, central force the work done in an infinitesimal deformation is
\begin{equation}
    \sum_i \mathbf{F}_i \cdot d \br_i = - \sum_{a=1}^{3} T_a \, d(\delta r_a),
    \qquad
    T_a \equiv f(\delta r_a; \delta r_b, \delta r_c),
\end{equation}
where $\{b, c\}$ are the two particles other than $a$, and $f$ is symmetric in its last two slots.
A potential exists if and only if the covector field $T = (T_1, T_2, T_3)$ on the three-dimensional shape space is conservative, i.e. its curl vanishes, $\varepsilon_{ijk} \partial_j T_k  = 0$, with $\varepsilon_{ij}$ the Levi-Civita tensor and $\partial_i \equiv \partial / \partial \delta r_i$.
Writing $f_2$ for the derivative of $f$ with respect to its second slot, and using the slot-$(2,3)$ symmetry to write $T_2 = f(\delta r_2; \delta r_1, \delta r_3)$, this becomes
\begin{equation}
    \label{eq:integrability}
    \dfrac{\partial f(x; y, z)}{\partial y} = \dfrac{\partial f(y; x, z)}{\partial x},
\end{equation}
i.e. the derivative of $f$ with respect to a \emph{spectator} bond must be unchanged when the acted bond and that spectator are exchanged.
The other two components of the curl give the same condition by relabeling.

Two remarks are in order regarding Eq.~\eqref{eq:integrability}.
First, why does the same condition not arise for pair forces? For conservative pair interactions, $f(x,y)=f(x-y)\equiv f(r)$ depends on a single scalar variable, so an antiderivative exists.
For three-body interactions, by contrast, no such guarantee applies.

Second, Eq.~\eqref{eq:integrability} is equivalent to requiring the total force to yield a symmetric dynamical matrix.
In Cartesian coordinates, it can be written as
$\partial F_i^\alpha / \partial r_j^\beta = D_{ij}^{\alpha\beta} = D_{ji}^{\beta\alpha} = \partial F_j^\beta / \partial r_i^\alpha$.

We emphasize that Eq.~\eqref{eq:integrability} constrains derivatives of $f$, not $f$ itself.
In particular, it does not require symmetry under $j\leftrightarrow k$.
The three-body term $a_{ijk} = a_3(\delta r_{ik}+\delta r_{jk})$ introduced below is a case in point: although $a_{ijk} \neq a_{ikj}$, it has constant $\partial_y f(x,y,z)=a_3$, so Eq.~\eqref{eq:integrability} holds and the force is conservative, with potential $U_3=a_3(\delta r_1\delta r_2+\delta r_1\delta r_3+\delta r_2\delta r_3)$.
As we show in Sec.~\ref{sec:taylor}, every term that is first order in $\delta r$ has constant $\partial_y f(x,y,z)$, so the minimal model is automatically integrable.
Violations first appear at second order, e.g.\ when $f \propto \delta r_{ij}(\delta r_{ik}+\delta r_{jk})$.
Such a force is reciprocal and exerts no net force or torque on a trimer, yet it can extract work from a cyclic deformation in shape space.

\textbf{Linear momentum.} Linear momentum is conserved if, for any configuration of particles, $\sum_i \mathbf{F}_i = 0$.
Considering a system with only 3 particles, an explicit calculation of the total force gives
\begin{equation}
    \label{eq:net_force}
    \mathbf{F}_\mathrm{net} = \sum_{i=1}^3 \mathbf{F}_i = 2 \sum_{\lbrace \mathrm{cyc} \rbrace} \left( a_{ijk}^- \hat{\mathbf{r}}_{ij} + b_{ijk}^- \hat{\mathbf{r}}_{ij}^\perp \right),
\end{equation}
where the sum is done over the cyclic set of three elements, $\lbrace \mathrm{cyc} \rbrace = \lbrace 123, 231, 312 \rbrace$.

\textbf{Angular momentum.} Similarly to linear momentum, a force law conserves angular momentum if, for any configuration, the torque vanishes, $\sum_i \mathbf{z}_i \times \mathbf{F}_i = 0$, where $\mathbf{z}_i$ is the displacement vector from a reference position $\bx$ to the position of particle $i$, $\mathbf{z}_i = \br_i - \bx$.
If $\mathbf{F}_\mathrm{net} \neq 0$, this will depend on the choice of $\bx$.
To avoid this ambiguity, we will use the centroid as the reference point, $\langle \br \rangle = \sum_{i=1}^N \br_i / N$, and write the displacement of each particle from the centroid as $\mathbf{z}_i = \br_i - \langle \br \rangle$.
Performing an explicit calculation of the net torque with respect to the centroid gives
\begin{equation}
    \label{eq:net_torque}
    \tau_\mathrm{net} =
    \sum_{i=1}^3 \mathbf{z}_i \times \mathbf{F}_i =
    \sum_{\lbrace \mathrm{cyc} \rbrace} \left[ \dfrac{4 \mathcal{A}}{3} \dfrac{a_{ijk}^-}{r_{ij}} - b_{ijk}^+ r_{ij} + b_{ijk}^- \dfrac{z_j^2 - z_i^2}{r_{ij}} \right],
\end{equation}
where $\mathcal{A} = (\br_{12} \times \br_{13})/2$ is the signed area of the triangle in question.
To get Eq.~\eqref{eq:net_torque}, we used $\mathbf{z}_i \times \hat{\br}_{ij} = (\mathbf{z}_i \times \mathbf{z}_j) / r_{ij}$, where the $r_{ij}$ in the denominator comes from the normalization of $\hat{\br}_{ij}$, and $\mathbf{z}_i \times \hat{\br}^\perp_{ij} = (\mathbf{z}_i \cdot \mathbf{z}_j - \vert \mathbf{z}_i \vert^2) / r_{ij}$.

The mismatched powers of $r_{ij}$ are an artifact of how the lever arms are written, not a genuine difference in distance dependence.
Every single-particle contribution begins with a $1/r_{ij}$ from normalizing the bond direction; the asymmetry appears only on pairing $(ijk)$ with $(jik)$.
The longitudinal lever arm $\mathbf{z}_i \times \mathbf{z}_j$ is antisymmetric under that exchange, so $a^+$ cancels and the $1/r_{ij}$ passes through untouched.
The transverse lever arm is neither symmetric nor antisymmetric, and its symmetric part is $-\tfrac{1}{2}|\mathbf{z}_i - \mathbf{z}_j|^2 = -\tfrac{1}{2} r_{ij}^2$, which cancels the $1/r_{ij}$ and leaves $-b^+_{ijk} r_{ij}$.
Mechanically, there are only two ways to make a torque: a net force applied somewhere, with lever arm the perpendicular distance from the centroid to the bond line; and a \emph{couple}, two equal and opposite transverse forces at the ends of a bond, whose moment is (force) $\times$ (separation) $= b^+_{ijk} r_{ij}$ and is independent of the reference point.
Equal and opposite forces \emph{along} a bond are collinear with their own line of action, so they generate no couple --- which is why $a^+$ is absent from Eq.~\eqref{eq:net_torque} altogether, and why the linear-in-$r_{ij}$ term lives exclusively in the chiral sector.

\subsection{Taylor expansion}
\label{sec:taylor}
To get a minimal model, we perform a Taylor expansion about $\delta r = 0$ to first order.
This gives
\begin{subequations}
\label{eq:coeffs}
    \begin{align}
        f(\delta r_{ij}; \delta r_{ik}, \delta r_{jk}) &= a_0 + k_a \, \delta r_{ij} + a_3 (\delta r_{ik} + \delta r_{jk}) + \alpha_3 (\delta r_{ik} - \delta r_{jk}) + O(\delta r^2), \\
        g(\delta r_{ij}; \delta r_{ik}, \delta r_{jk}) &= b_0 + k_b \, \delta r_{ij} + b_3 (\delta r_{ik} + \delta r_{jk}) + \beta_3 (\delta r_{ik} - \delta r_{jk}) + O(\delta r^2).
    \end{align}
\end{subequations}
Our minimal model is Eq.~\eqref{eq:eom_general} with coefficients given by Eq.~\eqref{eq:coeffs}.
Each coefficient has a direct interpretation.
Writing $\partial_{ij} f \equiv \partial f/\partial \delta r_{ij}$, we have:
\begin{itemize}
    \item $a_0 = f(0,0,0)$ is a uniform pre-stress on bond $ij$, which shifts the force-free bond length away from the reference spacing $\ell$.
    Likewise, $b_0 = g(0,0,0)$ is a pre-torque on bond $ij$; unlike $a_0$, it cannot be absorbed into a redefinition of $\ell$ because no bond length cancels a transverse force.

    \item $k_a = (\partial_{ij} f)|_0$ is a harmonic (pairwise) spring constant, giving $\mathbf{F}_i = k_a(r_{ij}-\ell)\hat{\br}_{ij}$.
    Similarly, $k_b = (\partial_{ij} g)|_0$ is a transverse spring constant, giving $\mathbf{F}_i = k_b(r_{ij}-\ell)\hat{\br}_{ij}^\perp$, which can generate odd elasticity.

    \item $a_3 = (\partial_{ik} f + \partial_{jk} f)/2$ is a reciprocal, conservative three-body term with potential $U=a_3(\delta r_1\delta r_2+\delta r_1\delta r_3+\delta r_2\delta r_3)$; similarly for $b_3$.

    \item $\alpha_3 = (\partial_{ik} f - \partial_{jk} f)/2$ violates linear momentum conservation. It measures the imbalance between the $ik$ and $jk$ bond deviations and is odd under $i\leftrightarrow j$.
    Similarly, $\beta_3$ is the chiral analog of $\alpha_3$.
\end{itemize}
Every coefficient in Eq.~\eqref{eq:coeffs} has constant $f_2$ (and likewise $g_2$), so Eq.~\eqref{eq:integrability} is satisfied identically.
Accordingly, the minimal model conserves energy whenever $\alpha_3 = \beta_3 = 0$ and $b_0 = k_b = b_3 = 0$.

\subsection{Propulsion}
We have $a_{ijk}^- = \alpha_3 (\delta r_{ik} - \delta r_{jk}) = \alpha_3 (\delta r_j - \delta r_i)$, where we have used the single-index notation introduced at the beginning of this section.
A similar result holds for $b_{ijk}^-$.
Then, the net force on a triplet of particles is given by
\begin{equation}
    \label{eq:trimer_netforce}
    \mathbf{F}_\mathrm{net} = 2 \alpha_3 \left(
        (\delta r_2 - \delta r_1) \hat{\br}_{12} +
        (\delta r_3 - \delta r_2) \hat{\br}_{23} +
        (\delta r_1 - \delta r_3) \hat{\br}_{31}
    \right) +
        \text{chiral terms}
\end{equation}
This expression is exact for all $\delta r_i$ (assuming Eq.~\eqref{eq:coeffs} holds).
When we consider small deviations from an equilateral triangle, we can get a simple expression if we write our vectors in complex notation, $(x, y) \to x+iy$.
Labeling the particles counter-clockwise, aligning $\hat{\br}_{12} = \hat{x}$, and defining $\zeta = e^{2\pi i / 3}$, we define the triangular shape parameter
\begin{equation} 
    Q = \delta r_1 + \zeta \, \delta r_2 + \zeta^2 \delta r_3.
\end{equation}
For an equilateral triangle at the reference spacing, $\delta r_i = 0$, so $Q_\mathrm{eq} = 0$.
The shape parameter $Q$ quantifies the shear of a triangle: $|Q|$ sets the shear magnitude and $\arg(Q)$ its orientation.
This parameterization is equivalent to taking a tangent plane at the north pole of Kendall's sphere~\cite{Kendall1999,Small2012}.
See~\cite{Klingenberg2020} for a modern review of statistical shape theory in morphometrics.

We will replace the bond vectors $\hat{\br}_{ij}$ by their equilateral-triangle counterparts, which introduces errors of order $\delta r^2$ (which we have already dropped in defining the functions $f, g$).
With the labeling above, the three bond directions sit at $0$, $2\pi/3$ and $4\pi/3$, so
\begin{equation}
    (\hat{\br}_{12}, \hat{\br}_{23}, \hat{\br}_{31}) \to (1, \zeta, \zeta^2).
\end{equation}
With this replacement, the net force is then given by (focusing only on the term proportional to $\alpha_3$)
\begin{align*}
    \mathbf{F}_\mathrm{net}^a &= 2\alpha_3 \left[
        \delta r_1 (\hat{\br}_{31} - \hat{\br}_{12}) +
        \delta r_2 (\hat{\br}_{12} - \hat{\br}_{23}) +
        \delta r_3 (\hat{\br}_{23} - \hat{\br}_{31})
    \right] \\
    &= 2\alpha_3 \left[
        \delta r_1 (\zeta^2 - 1) +
        \delta r_2 (1 - \zeta) +
        \delta r_3 (\zeta - \zeta^2)
    \right] \\
    &= 2\alpha_3 (\zeta^2-1) \left[
        \delta r_1 +
        \delta r_2 \dfrac{1 - \zeta}{\zeta^2 - 1} +
        \delta r_3 \dfrac{\zeta(1 - \zeta)}{\zeta^2 - 1}
    \right] \\
    &= 2\alpha_3 (\zeta^2-1) \left[
        \delta r_1 +
        \delta r_2 \zeta^{-2} +
        \delta r_3 \zeta^{-1}
    \right] \\
    &= 2\alpha_3 (\zeta^2-1) \left[
        \delta r_1 +
        \delta r_2 \zeta +
        \delta r_3 \zeta^2
    \right] \\
    &= 2\alpha_3 (\zeta^2-1) Q,
\end{align*}
where we used $1 + \zeta + \zeta^2 = 0$ to write $\zeta^2 - 1 = (\zeta+1)(\zeta-1) = \zeta^2 (1-\zeta)$ and $\zeta^n = \zeta^{n+3}$.
The force coming from the chiral forces is recovered by replacing every $\hat{\br}_{ij} \to \hat{\br}^\perp_{ij} = e^{i\pi/2}\hat{\br}_{ij}$.
The total net force is then
\begin{equation}
    \label{eq:netForce_linear}
    \mathbf{F}_\mathrm{net} = 2(\zeta^2-1) Q \left( \alpha_3 + e^{i\pi / 2} \beta_3 \right).
\end{equation}
Propulsion is thus linear in the shear $Q$ and vanishes on the equilateral triangle of any size.

\subsection{Harmonic theory}
\begin{wrapfigure}{r}{0.5\textwidth}
    \centering
    \vspace{-0.75\baselineskip}
    \includegraphics[width=0.48\textwidth]{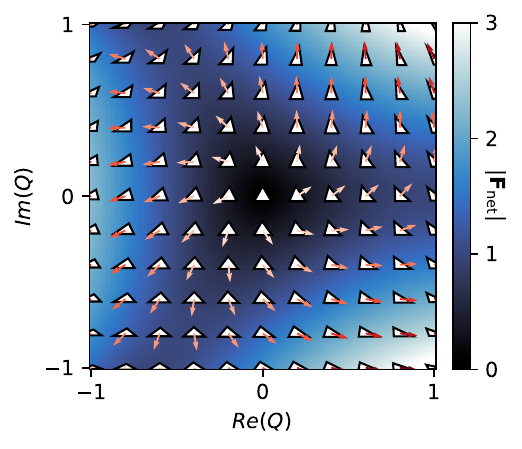}
    \caption{\textbf{Net force in shape space.} Net force, Eq.~\eqref{eq:trimer_netforce}, given in $Q$-space for $k_a = -\alpha_3 = 1$ and all other coefficients set to zero. Colors give the magnitude of the net force, arrows give the direction (arrow colors are chosen for visualization purposes and do not carry physical meaning). The triangles show the configuration in $Q$-space.}
    \label{fig:propulsion}
    \vspace{-0.75\baselineskip}
\end{wrapfigure}
Suppose each particle sits at a reference position $\bR_i$ and its instantaneous position is $\br_i = \bR_i + \mathbf{u}_i$, with $\mathbf{u}_i$ a small displacement.
The relative displacement on bond $(i,j)$ is $\mathbf{u}_{ij} = \mathbf{u}_j - \mathbf{u}_i$.
To linear order in $u_{ij}/\ell$ (with reference spacing $\ell = R_{ij}$), the force separates into a term from variations of the scalar coefficients and a geometric term from bond rotation:
\begin{equation}
    \mathbf{F}_i^\mathrm{lin} = \sum_{(j,k)} \delta a_{ijk}\,\hat{\bR}_{ij} + a_{ijk}^0\,\delta\hat{\br}_{ij} + \text{chiral terms}.
\end{equation}
To evaluate $\delta a_{ijk}$ we first linearize the bond-length deviation,
\begin{align*}
    \delta r_{ij} &= |\br_j-\br_i| - \ell = |\mathbf{R}_{ij}+\mathbf{u}_{ij}| - \ell \\
    &\approx \mathbf{u}_{ij}\cdot\hat{\mathbf{R}}_{ij} \equiv u_{ij}^\parallel,
\end{align*}
so to first order $\delta r_{ij}$ is simply the projection of $\mathbf{u}_{ij}$ along the reference bond direction $\hat{\mathbf{R}}_{ij}$.
Transverse motion affects the bond length only at second order.
We therefore obtain
\begin{align}
    \delta a_{ijk} &=
    k_a u_{ij}^\parallel +
    a_3 (u_{ik}^\parallel + u_{jk}^\parallel) +
    \alpha_3 (u_{ik}^\parallel - u_{jk}^\parallel) \\
    \delta b_{ijk} &=
    k_b u_{ij}^\parallel +
    b_3 (u_{ik}^\parallel + u_{jk}^\parallel) +
    \beta_3 (u_{ik}^\parallel - u_{jk}^\parallel).
\end{align}
The pre-stress and pre-torque terms multiply $\delta \hat{\br}_{ij}$ and $\delta \hat{\br}^\perp_{ij}$ respectively, where
\begin{align}
    \delta \hat{\br}_{ij} &= \dfrac{\hat{\bR}_{ij}^\perp \cdot \mathbf{u}_{ij}}{\ell}\hat{\bR}_{ij}^\perp =
    \dfrac{u_{ij}^\perp}{\ell} \hat{\bR}_{ij}^\perp; \;
    \delta \hat{\br}_{ij}^\perp = -\dfrac{u_{ij}^\perp}{\ell} \hat{\bR}_{ij}.
\end{align}
The second follows from the first by applying $R^{\pi/2}$ and using $R^{\pi/2} \hat{\bR}^\perp = -\hat{\bR}$.
Note that a unit vector is insensitive to motion along itself: stretching scales $\br_{ij}$ and $|\br_{ij}|$ by the same factor, so only $u^\perp_{ij}$ rotates the bond.
Our total force is therefore given by
\begin{equation}
\label{eq:Flin}
\begin{split}
    \mathbf{F}_i^\mathrm{lin} = \sum_{(j, k)} \bigg[ &
    \left( k_a u_{ij}^\parallel + a_3 (u_{ik}^\parallel + u_{jk}^\parallel) + \alpha_3 (u_{ik}^\parallel - u_{jk}^\parallel) \right) \hat{\bR}_{ij} +
    a_{ijk}^0  \dfrac{u_{ij}^\perp}{\ell} \hat{\bR}_{ij}^\perp \\
    +& \left( k_b u_{ij}^\parallel + b_3 (u_{ik}^\parallel + u_{jk}^\parallel) + \beta_3 (u_{ik}^\parallel - u_{jk}^\parallel) \right) \hat{\bR}_{ij}^\perp -
    b_{ijk}^0  \dfrac{u_{ij}^\perp}{\ell} \hat{\bR}_{ij} \bigg].
\end{split}
\end{equation}
On an unstrained lattice $a^0_{ijk} = a_0$ and $b^0_{ijk} = b_0$.
It is worth noting the structure of the two pre-stress terms: $a_0$ produces a transverse force from transverse motion, similar to Euler buckling, whereas $b_0$ produces a longitudinal force from transverse motion.

\subsection{Dynamical matrix and elastic moduli}
    \subsubsection{Setup and conventions}
        We now place the particles on a triangular lattice of spacing $\ell$ and ask what elastic moduli the minimal model produces.
        The dynamical matrix and its Fourier transform are defined by
        \begin{equation}
            \label{eq:dynmat_def_supp}
            F^\alpha(\bx) = - \sum_{\by} D^{\alpha\beta}(\bx - \by) \, u^\beta(\by),
            \qquad
            D^{\alpha\beta}(\bq) = \sum_{\bx} e^{-i \bq \cdot \bx} D^{\alpha\beta}(\bx),
        \end{equation}
        so that $F^\alpha(\bq) = - D^{\alpha\beta}(\bq) u^\beta(\bq)$.
        We take the interaction to act only on nearest-neighbor triangles.
        
        On a triangular lattice, each site belongs to six triangles, and each triangle contributes two ordered pairs $(j,k)$; accordingly, the sum $\sum_{(j,k)}$ in Eq.~\eqref{eq:eom_general} runs over twelve terms per site (Fig.~\ref{fig:force_sum}b).
        One could instead sum over all pairs of a site's six neighbors, giving $\binom{6}{2}=15$ unordered (or $30$ ordered) terms.
        However, that rule is not permutation-closed.
        If $j$ and $k$ sit on opposite sides of $i$, then $|\br_{jk}| = 2\ell$, so from $j$'s point of view $k$ is not a neighbor and the triple $\{i,j,k\}$ never appears in $\mathbf{F}_j$.
        The selection rule must be a property of the triple, not of a single vertex. Otherwise Eq.~\eqref{eq:net_force} fails and even a purely reciprocal force law ($\alpha_3=\beta_3=0$) acquires a spurious net force.
        
        \subsubsection{Fourier transforming the linearized force}
        
        We insert a plane wave $\mathbf{u}_m = \mathbf{U} e^{i \bq \cdot \bR_m}$ into Eq.~\eqref{eq:Flin} and use Bloch's theorem to replace relative displacements with phase differences.
        Writing $E_j \equiv e^{i \bq \cdot \bR_{ij}}$ and $E_k \equiv e^{i\bq\cdot\bR_{ik}}$,
        \begin{equation}
            \label{eq:phases}
            u_{ij}^\parallel \to (E_j - 1)\, \mathbf{U}\cdot\hat{\bR}_{ij},
            \qquad
            u_{ik}^\parallel \to (E_k - 1)\, \mathbf{U}\cdot\hat{\bR}_{ik},
            \qquad
            u_{jk}^\parallel \to (E_k - E_j)\, \mathbf{U}\cdot\hat{\bR}_{jk},
        \end{equation}
        and likewise $u_{ij}^\perp \to (E_j - 1)\, \mathbf{U}\cdot\hat{\bR}_{ij}^\perp$.
        Unlike the analogous pairwise calculation, the three bonds of a triangle carry three different phase factors, introducing the three-body structure to the calculation.
        
        It is convenient to name the two independent projections of $\bq$ on the triangle,
        \begin{equation}
            \phi_j \equiv \bq \cdot \hat{\bR}_{ij},
            \qquad
            \phi_k \equiv \bq \cdot \hat{\bR}_{ik},
            \qquad\text{so that}\qquad
            \bq \cdot \hat{\bR}_{jk} = \phi_k - \phi_j,
        \end{equation}
        where we got the last identity by using $\hat{\bR}_{jk} = \hat{\bR}_{ik} - \hat{\bR}_{ij}$.
        Expanding the phases,
        \begin{equation}
            E_j - 1 = i \ell \phi_j - \tfrac{1}{2} \ell^2 \phi_j^2 + \dots,
            \qquad
            E_k - E_j = i \ell (\phi_k - \phi_j) - \tfrac{1}{2} \ell^2 (\phi_k^2 - \phi_j^2) + \dots .
        \end{equation}
        Two orders drop out before we reach the elastic term.
        The $O(q^0)$ piece vanishes identically because only differences of displacements appear, giving the acoustic sum rule, $\sum_{\bx} D^{\alpha\beta}(\bx) = 0$ (equivalent to translation invariance).
        The $O(q^1)$ piece vanishes by inversion symmetry: the twelve ordered pairs come in inverted partners $(\bR_{ij}, \bR_{ik}) \to (-\bR_{ij}, -\bR_{ik})$, under which every $\phi$ changes sign while every dyad $\hat{R}_{ab}^\alpha \hat{R}_{cd}^\beta$ is even.
        The first surviving term is therefore $O(q^2)$.

    \subsubsection{The dynamical matrix}
    
        Collecting the $O(q^2)$ terms and using $D^{\alpha\beta} = -\partial F^\alpha/\partial U^\beta$,
        \begin{equation}
            \label{eq:dynmat_moments}
            \begin{split}
            D^{\alpha\beta}(\bq) = \frac{\ell^2}{2} \sum_{(j,k)} \bigg[
            &\; k_a\, \hat{R}_{ij}^\alpha \hat{R}_{ij}^\beta \hat{R}_{ij}^\gamma \hat{R}_{ij}^\eta
             + (a_3 + \alpha_3)\, \hat{R}_{ij}^\alpha \hat{R}_{ik}^\beta \hat{R}_{ik}^\gamma \hat{R}_{ik}^\eta
            + (a_3 - \alpha_3) \hat{R}_{ij}^\alpha \hat{R}_{jk}^\beta \big( \hat{R}_{ik}^\gamma \hat{R}_{ik}^\eta - \hat{R}_{ij}^\gamma \hat{R}_{ij}^\eta \big) \\
            &+ k_b\, \hat{R}_{ij}^{\perp\alpha} \hat{R}_{ij}^\beta \hat{R}_{ij}^\gamma \hat{R}_{ij}^\eta
             + (b_3 + \beta_3)\, \hat{R}_{ij}^{\perp\alpha} \hat{R}_{ik}^\beta \hat{R}_{ik}^\gamma \hat{R}_{ik}^\eta
            + (b_3 - \beta_3) \hat{R}_{ij}^{\perp\alpha} \hat{R}_{jk}^\beta \big( \hat{R}_{ik}^\gamma \hat{R}_{ik}^\eta - \hat{R}_{ij}^\gamma \hat{R}_{ij}^\eta \big)\\
            &+ \frac{1}{\ell} \Big( a_0\, \hat{R}_{ij}^{\perp\alpha} - b_0\, \hat{R}_{ij}^{\alpha} \Big) \hat{R}_{ij}^{\perp\beta} \hat{R}_{ij}^\gamma \hat{R}_{ij}^\eta
            \bigg] \; q^\gamma q^\eta
            \;+\; O(q^4).
            \end{split}
        \end{equation}
        The lines are, in order: the parity-even stiffnesses, the chiral stiffnesses, and the two pre-stresses.
        Note that the chiral lines are the parity-even lines with the replacement $\hat{R}_{ij}^\alpha \to \hat{R}_{ij}^{\perp\alpha}$ in the first index of $D^{\alpha \beta}$, which gives the direction of the force.
        The pre-stress line carries an extra $1/\ell$ because $a_0$ and $b_0$ carry units of force, while the dynamical matrix has units of force/length.
        
        Eq.~\eqref{eq:dynmat_moments} is the three-body generalization of a familiar result.
        For a simple central spring, the whole dynamical matrix is fixed by the fourth moment of the bond directions, $\hat{n}^\alpha \hat{n}^\beta \hat{n}^\gamma \hat{n}^\eta$, in which all four factors refer to the same bond~\cite{Born1954}.
        Here, the four factors in each summand can refer to different bonds of the triangle.
        
        Evaluating Eq.~\eqref{eq:dynmat_moments} therefore requires the fourth moments summed over the twelve ordered pairs.
        Writing $\Delta^{\alpha\beta\gamma\eta} \equiv \delta^{\alpha\beta}\delta^{\gamma\eta} + \delta^{\alpha\gamma}\delta^{\beta\eta} + \delta^{\alpha\eta}\delta^{\beta\gamma}$ for the isotropic combination, three of the four are simply proportional to it:
        \begin{subequations}
        \label{eq:moments4}
        \begin{align}
            \sum_{(j,k)} \hat{R}_{ij}^\alpha \hat{R}_{ij}^\beta \hat{R}_{ij}^\gamma \hat{R}_{ij}^\eta &= \tfrac{3}{2}\, \Delta^{\alpha\beta\gamma\eta}, \\
            \sum_{(j,k)} \hat{R}_{ij}^\alpha \hat{R}_{ik}^\beta \hat{R}_{ik}^\gamma \hat{R}_{ik}^\eta &= \tfrac{3}{4}\, \Delta^{\alpha\beta\gamma\eta}, \\
            \sum_{(j,k)} \hat{R}_{ij}^\alpha \hat{R}_{jk}^\beta \hat{R}_{ij}^\gamma \hat{R}_{ij}^\eta &= -\tfrac{3}{4}\, \Delta^{\alpha\beta\gamma\eta}, \\
            \sum_{(j,k)} \hat{R}_{ij}^\alpha \hat{R}_{jk}^\beta \hat{R}_{ik}^\gamma \hat{R}_{ik}^\eta &= -3\,\delta^{\alpha\beta}\delta^{\gamma\eta} + \tfrac{3}{2}\big( \delta^{\alpha\gamma}\delta^{\beta\eta} + \delta^{\alpha\eta}\delta^{\beta\gamma} \big),
        \end{align}
        \end{subequations}
        with the corresponding moments carrying a $\perp$ on the first index obtained by $\delta^{\alpha\bullet} \to \varepsilon^{\alpha\bullet}$ in that slot.
        The second moments are
        \begin{equation}
            \label{eq:moments}
            \sum_{(j,k)} \hat{R}_{ij}^\alpha \hat{R}_{ij}^\beta = 6\, \delta^{\alpha \beta},
            \qquad
            \sum_{(j,k)} \hat{R}_{ij}^\alpha \hat{R}_{ik}^\beta = 3\, \delta^{\alpha \beta},
            \qquad
            \sum_{(j,k)} \hat{R}_{ij}^\alpha \hat{R}_{jk}^\beta = -3\, \delta^{\alpha \beta}.
        \end{equation}
        The isotropy in both cases follows from the neighbor shell being six-fold symmetric.
        Only the last line of Eq.~\eqref{eq:moments4} is genuinely new, and it is the one that generates the $(a_3 - \alpha_3)$ block below.

    \subsubsection{Elastic moduli for three-body force}
    
        Resolve the displacement into longitudinal and transverse parts, $u^\parallel = \hat{\bq}\cdot\mathbf{u}$ and $u^\perp = \hat{\bq}^\perp\cdot\mathbf{u}$, and define the rotated, normalized dynamical matrix
        \begin{equation}
            \label{eq:M_def}
            \tilde{D} \equiv \frac{1}{V q^2}\, \mathsf{R}^{\!\top} D(\bq)\, \mathsf{R},
            \qquad
            \mathsf{R} = \big[\, \hat{\bq} \;\; \hat{\bq}^\perp \,\big],
        \end{equation}
        where $V = \tfrac{\sqrt{3}}{2} \ell^2$ is the area of the real-space unit cell.
        For an isotropic two-dimensional odd solid $\tilde{D}$ is independent of the direction of $\bq$ and takes the form
        \begin{equation}
            \label{eq:modulus_matrix}
            \tilde{D} = \begin{pmatrix}
                B + G & K^{o} \\
                A^{o} - K^{o} & G
            \end{pmatrix}.
        \end{equation}
        The moduli can then be read off directly from this matrix as $G = \tilde{D}_{22}$, $B = \tilde{D}_{11} - \tilde{D}_{22}$, $K^{o} = \tilde{D}_{12}$ and $A^{o} = \tilde{D}_{21} + \tilde{D}_{12}$.
        This is worth doing rather than diagonalizing: the eigenvalues of $D(\bq)$ do not label the branches, since the longitudinal branch is softer than the transverse one over part of parameter space, and once $K^{o} \neq 0$ the matrix is non-normal and its eigenvalues become complex.
        
        Substituting Eq.~\eqref{eq:moments4} into Eq.~\eqref{eq:dynmat_moments} and rotating into the $(\hat{\bq}, \hat{\bq}^\perp)$ frame, every contribution collapses onto a fixed, direction-independent $2\times 2$ block with integer entries:
        \begin{equation}
            \label{eq:blocks}
            \begin{split}
            \tilde{D} = \frac{\sqrt{3}}{4} \bigg[
            &\; k_a \begin{pmatrix} 6 & 0 \\ 0 & 2 \end{pmatrix}
            + (a_3 + \alpha_3) \begin{pmatrix} 3 & 0 \\ 0 & 1 \end{pmatrix}
            + (a_3 - \alpha_3) \begin{pmatrix} 3 & 0 \\ 0 & -3 \end{pmatrix}
            + \frac{a_0}{\ell} \begin{pmatrix} 2 & 0 \\ 0 & 6 \end{pmatrix} \\[4pt]
            &+ k_b \begin{pmatrix} 0 & -2 \\ 6 & 0 \end{pmatrix}
            + (b_3 + \beta_3) \begin{pmatrix} 0 & -1 \\ 3 & 0 \end{pmatrix}
            + (b_3 - \beta_3) \begin{pmatrix} 0 & 3 \\ 3 & 0 \end{pmatrix}
            + \frac{b_0}{\ell} \begin{pmatrix} 0 & -6 \\ 2 & 0 \end{pmatrix}
            \bigg].
            \end{split}
        \end{equation}
        As a worked example, take the $k_a$ term.
        Nothing in it depends on the spectator particle, so the twelve ordered pairs reduce to twice a sum over the six neighbors.
        The first line of Eq.~\eqref{eq:moments4} then gives $\tfrac{3}{2}\Delta^{\alpha\beta\gamma\eta} q^\gamma q^\eta = \tfrac{3}{2}(\delta^{\alpha\beta} q^2 + 2 q^\alpha q^\beta)$, so $D^{\alpha\beta}|_{k_a} = \tfrac{3}{4} \ell^2 k_a (\delta^{\alpha\beta} q^2 + 2 q^\alpha q^\beta)$.
        In the $(\hat{\bq}, \hat{\bq}^\perp)$ frame $q^\alpha q^\beta / q^2 \to \mathrm{diag}(1,0)$, and dividing by $V q^2 = \tfrac{\sqrt{3}}{2} \ell^2 q^2$ the factors of $\ell^2$ cancel, leaving $\tfrac{\sqrt{3}}{2} k_a \,\mathrm{diag}(3,1)$.
        The $3\!:\!1$ ratio is nothing but the $\delta^{\alpha\beta} q^2 + 2 q^\alpha q^\beta$ structure of the isotropic fourth moment: the longitudinal direction receives $1+2$, the transverse $1+0$.
        
        Comparing Eq.~\eqref{eq:blocks} with Eq.~\eqref{eq:modulus_matrix} gives the four moduli,
        \begin{subequations}
        \label{eq:moduli_supp}
            \begin{align}
                B &= \sqrt{3}\left( k_a + 2 a_3 - \alpha_3 - \frac{a_0}{\ell} \right),
                &G = \frac{\sqrt{3}}{2}\left( k_a - a_3 + 2 \alpha_3 + \frac{3 a_0}{\ell} \right), \\
                A^{o} &= \sqrt{3}\left( k_b + 2 b_3 - \beta_3 - \frac{b_0}{\ell} \right),
                &K^{o} = \frac{\sqrt{3}}{2}\left( -k_b + b_3 - 2 \beta_3 - \frac{3 b_0}{\ell} \right).
            \end{align}
        \end{subequations}
        The factor $1/\ell$ appears only on $a_0$ and $b_0$: those are pre-stresses, with units of force, and it takes a length to convert a force into a two-dimensional modulus.
        
        Some remarks:
        \paragraph{The three-body structure is invisible in the bulk.} Eight coefficients
        enter through eight blocks, but those blocks span only a four-dimensional space.
        For example, the $(a_3 + \alpha_3)$ block is exactly half the $k_a$ block.
        Only four combinations survive, giving our four moduli.
        The collapse happens at the level of the moments, Eq.~\eqref{eq:moments4}, not in the force law.
        Three of the four mixed moments come out proportional to the single-bond moment, with the constants of proportionality being the cosines $\pm\tfrac{1}{2}$ of an equilateral triangle.
        On a non-Bravais lattice, such as a Kagome lattice, the shell is no longer six-fold symmetric and this proportionality should fail.
        \paragraph{Consequences for measurement.}
        Because only four combinations survive, a phonon measurement on a defect-free raft cannot separate $\alpha_3$ from the pre-stress $a_0$; both renormalize $B$ and $G$, differing only in the ratio $\delta G/\delta B$, which is $-1$ for $\alpha_3$ and $-3/2$ for $a_0$.
        Separating them cleanly requires an isolated trimer or a defect.
        Conversely, $b_0$ and $k_b$ are separable, since they enter $A^{o}$ and $K^{o}$ in different ratios.

\subsection{Elastic moduli for general linearized forces}
    A particle at rest position $\bx$ interacts with a neighbor at rest position $\bx' = \bx + \lattvec{n}$, where $\lattvec{n}$ is a lattice vector of length $\ell$.
    Under a small deformation, the particles move to $\bx + \mathbf{u}(\bx)$ and $\bx' + \mathbf{u}(\bx')$.
    The force depends on their separation vector
    \begin{equation}
        \Delta\bx \equiv \big(\bx' + \mathbf{u}(\bx')\big) - \big(\bx + \mathbf{u}(\bx)\big)
        = \lattvec{n} + \delta\mathbf{u}(\lattvec{n}),
    \end{equation}
    where we define $\delta\mathbf{u}(\lattvec{n}) \equiv \mathbf{u}(\bx + \lattvec{n}) - \mathbf{u}(\bx)$.

    As discussed in the main text, linearizing any lattice force law yields a bond-frame relation of the form
    \begin{equation}
        \begin{pmatrix} F_r \\ F_\phi \end{pmatrix}
        =-
        \begin{pmatrix} k_{rr} & k_{r\phi} \\ k_{\phi r} & k_{\phi\phi} \end{pmatrix}
        \begin{pmatrix} \delta r \\ \delta\phi \end{pmatrix},
    \end{equation}
    where $\delta r = r-r_0$ and $\delta\phi = \phi-\phi_0$.

    In Cartesian coordinates, this can be written as
    \begin{equation}
    \begin{split}
        \mathbf{F}(\Delta\bx) ={}& -k_{rr}\,\frac{\Delta\bx}{|\Delta\bx|}\big(|\Delta\bx| - \ell\big)
        - k_{r\phi}\,\frac{\Delta\bx}{|\Delta\bx|}\cdot\big(\Delta\bx\times\ulattvec{n}\big) \\
        &- k_{\phi r}\,\frac{\Delta\bx}{|\Delta\bx|}\times\big(|\Delta\bx| - \ell\big)
        - k_{\phi\phi}\,\frac{\Delta\bx}{|\Delta\bx|}\times\big(\Delta\bx\times\ulattvec{n}\big),
    \end{split}
    \end{equation}
    using $|\Delta\bx\times\ulattvec{n}|\approx\delta\phi$ for small angles.
    The triple product is equivalent to $|\Delta\bx\times\ulattvec{n}|\,\hat{\boldsymbol{\phi}}$, since $\hat{\boldsymbol{\phi}}$ is orthogonal to $\Delta\bx/|\Delta\bx|$.

    Linearizing for $|\delta\mathbf{u}|/\ell \ll 1$ gives
    \begin{equation}
    \label{eq:effectiveForce}
        \mathbf{F}(\bx-\bx') =
        \big(k_{rr}\,\ulattvec{n} + k_{\phi r}\,\ulattvecperp{n}\big)\big(\delta\mathbf{u}(\lattvec{n})\cdot\ulattvec{n}\big)
        + \big(k_{r\phi}\,\ulattvec{n} + k_{\phi\phi}\,\ulattvecperp{n}\big)\big(\delta\mathbf{u}(\lattvec{n})\cdot\ulattvecperp{n}\big).
    \end{equation}
    Here $\lattvecperp{n}$ is perpendicular to $\lattvec{n}$, with components $\lattcompperp{n}{i} = \varepsilon_{ij}\lattcomp{n}{j}$.
    The term $\delta\mathbf{u}(\lattvec{n})\cdot\ulattvec{n}$ is the displacement along the bond direction (i.e. $\delta r$), whereas $\delta\mathbf{u}(\lattvec{n})\cdot\ulattvecperp{n}$ is the transverse displacement (i.e. $\delta r^\perp$).
    Coefficients multiplying $\ulattvec{n}$ produce forces along the bond, while those multiplying $\ulattvecperp{n}$ produce forces transverse to the bond.

    Given that the particle exists on a triangular lattice, the total force on the particle at $\bx$ is given by the sum of the forces from all 6 of its neighbors.
    The lattice vectors are
    \begin{equation}
    \label{eq:lattice_vectors}
        \lattvec{1} = \dfrac{\ell}{2} \begin{pmatrix} \sqrt{3} \\ 1 \end{pmatrix}; \quad 
        \lattvec{2} = \dfrac{\ell}{2} \begin{pmatrix} -\sqrt{3} \\ 1 \end{pmatrix}; \quad 
        \lattvec{3} = \ell \begin{pmatrix} 0 \\ -1 \end{pmatrix} ,
    \end{equation}
    with $\lattvecperp{n}$ defined accordingly.
    The total force is
    \begin{equation}
    \label{eq:totalForce}
    \begin{aligned}
        \mathbf{F}(\bx) = \sum_{n = 1}^3 &\mathbf{F} \left( \bx, \bx + \lattvec{n} \right) + \mathbf{F} \left( \bx, \bx - \lattvec{n} \right) \\
        = \sum_{n=1}^3 
        \bigg( 
        & \left( k_{rr} \ulattvec{n} + k_{\phi r} \ulattvecperp{n} \right)
            \left[ 
                \left( \delta \mathbf{u}(\lattvec{n}) + \delta\mathbf{u}(-\lattvec{n}) \right) \cdot \ulattvec{n} 
            \right] \\
        +& \left( k_{r \phi} \ulattvec{n} +  k_{\phi \phi} \ulattvecperp{n} \right)
            \left[
                \left( \delta \mathbf{u}(\lattvec{n}) + \delta\mathbf{u}(-\lattvec{n}) \right) \cdot \ulattvecperp{n} 
            \right]
        \bigg).
    \end{aligned}
    \end{equation}
    The total force is linear in the displacement $\delta \mathbf{u}$, allowing us to write
    \begin{equation}
        F_i(\bx) = -\sum_{\bx'} D_{ij}(\bx - \bx') \delta u_j(\bx'),
    \end{equation}
    where $D_{ij}(\bx)$ is the dynamical matrix,
    \begin{equation}
        \label{eq:dynmat}
        D_{ij}(\bx) = -\sum_{n=1}^3
        \left(
            k_{rr}\,\ulattcomp{n}{i}\ulattcomp{n}{j}
            + k_{\phi r}\,\ulattcompperp{n}{i}\ulattcomp{n}{j}
            + k_{r\phi}\,\ulattcomp{n}{i}\ulattcompperp{n}{j}
            + k_{\phi\phi}\,\ulattcompperp{n}{i}\ulattcompperp{n}{j}
        \right)
        \left[\delta(\bx+\lattvec{n}) - 2\delta(\bx) + \delta(\bx-\lattvec{n})\right].
    \end{equation}
    Here $\delta(\bx)$ is the Dirac delta function, and we used
    $\delta\mathbf{u}(\lattvec{n}) + \delta\mathbf{u}(-\lattvec{n}) = \mathbf{u}(\bx + \lattvec{n}) - 2\mathbf{u}(\bx) + \mathbf{u}(\bx - \lattvec{n})$.

    Defining the Fourier transform as $f(\bq)=\sum_{\bx}e^{-i\bq\cdot\bx}f(\bx)$, we obtain
    \begin{equation}
        \label{eq:dynmat_q}
        D_{ij}(\bq) = 2\sum_{n=1}^3
        \left(
            k_{rr}\,\ulattcomp{n}{i}\ulattcomp{n}{j}
            + k_{\phi r}\,\ulattcompperp{n}{i}\ulattcomp{n}{j}
            + k_{r\phi}\,\ulattcomp{n}{i}\ulattcompperp{n}{j}
            + k_{\phi\phi}\,\ulattcompperp{n}{i}\ulattcompperp{n}{j}
        \right)
        \left[1-\cos(\bq\cdot\lattvec{n})\right].
    \end{equation}
    Expanding at small $q=\|\bq\|$ gives
    \begin{equation}
        \label{eq:dynmat_smallq}
        D_{ij}(\bq) = \frac{3q^2\ell^2}{8}
        \begin{pmatrix}
            3k_{rr}+k_{\phi\phi} & 3k_{r\phi}-k_{\phi r} \\
            3k_{\phi r}-k_{r\phi} & k_{rr}+3k_{\phi\phi}
        \end{pmatrix}.
    \end{equation}

    A two-dimensional elastic material that is isotropic and does not couple to rotations generically has~\cite{Fruchart2023}
    \begin{equation}
        \label{eq:dynmat_generic}
        D_{ij}(\bq) = Vq^2
        \begin{pmatrix}
            B+G & K^o \\
            A^o-K^o & G
        \end{pmatrix},
    \end{equation}
    where $V=\sqrt{3}\ell^2/2$ is the unit-cell area, $B$ is the bulk modulus, $G$ is the shear modulus, $A^o$ couples dilation to torque, and $K^o$ is an odd shear modulus.
    Comparing Eq.~\eqref{eq:dynmat_generic} with Eq.~\eqref{eq:dynmat_smallq} yields
    \begin{equation}
        \label{eq:moduli_eff}
        \begin{aligned}
            B &= \frac{\sqrt{3}}{2}(k_{rr}-k_{\phi\phi});
            &\quad G &= \frac{\sqrt{3}}{4}(k_{rr}+3k_{\phi\phi}); \\
            A^o &= \frac{\sqrt{3}}{2}(k_{r\phi}+k_{\phi r});
            &\quad K^o &= \frac{\sqrt{3}}{4}(3k_{r\phi}-k_{\phi r}).
        \end{aligned}
    \end{equation}
    Together with Eq.~\eqref{eq:moduli_supp}, this gives the correspondence between the effective stiffnesses $k_{ij}$ and the minimal-model coefficients.

    We check this algebra against the implemented force law by computing $D$ via central differences on an interior site of a perfect triangular-lattice disk, displacing one particle at a time by $h=10^{-5}\ell$.
    This route uses only the force law of Eq.~(2) of the main text, independent of the derivation above.
    The resulting long-wavelength moduli agree with Eq.~\eqref{eq:moduli_supp} to relative error $6\times10^{-11}$ for all directions of $\bq$.
    This holds for each coefficient in isolation and for random parameter sets with all eight (including chiral) coefficients nonzero.
    The measured bond blocks, rotated into each bond frame, reproduce $k_{rr}$ and $k_{\phi\phi}$ of Eq.~(4) of the main text to the same precision.
    They also reproduce the chiral stiffnesses $k_{r\phi}$ and $k_{\phi r}$ that follow from Eqs.~\eqref{eq:moduli_supp} and~\eqref{eq:moduli_eff}.
    Because the three-body force couples only nearest neighbors on the lattice, Eq.~\eqref{eq:dynmat_q} holds at every wavevector.
    The measured $D(\bq)$ along $\Gamma$--K and $\Gamma$--M agrees with it to $5\times10^{-9}$, with the largest deviation at the smallest $q$, where finite differences limit the precision.

\section{Elastodynamics}
    The bulk equation of motion can be written as
    \begin{equation}
    \label{eq:elastodynamics_supp}
        \rho\,\partial_{tt}u_j + \gamma\,\partial_t u_j = F_j(\bx) = -\sum_{\bx'} D_{jn}(\bx-\bx')\,u_n(\bx').
    \end{equation}
    We seek plane-wave solutions propagating along $\bq$,
    $u_i(\bx,t)=u_i^0(\bq)\exp\!\left(i(\bq\cdot\bx-\omega t)\right)$.
    Substituting into Eq.~\eqref{eq:elastodynamics_supp} gives the eigenvalue problem
    \begin{equation}
        -(\rho\omega^2+i\gamma\omega)\,u_j^0 = -D_{jn}(\bq)\,u_n^0,
    \end{equation}
    where $D(\bq)$ is the dynamical matrix of Eq.~\eqref{eq:dynmat_q}.

    In the longitudinal--transverse basis, $u_r = \hat{\bq}\cdot\mathbf{u}$ and $u_\phi = \hat{\bq}\times\mathbf{u}$, the problem diagonalizes to
    \begin{equation}
        -(\rho\omega^2+i\gamma\omega)
        \begin{pmatrix} u_r \\ u_\phi \end{pmatrix}
        = -q^2
        \begin{pmatrix}
            B+G  & 0 \\
            0 & G
        \end{pmatrix}
        \begin{pmatrix} u_r \\ u_\phi \end{pmatrix}.
    \end{equation}
    For $\gamma=0$, the eigenfrequencies give the phase velocities $c=\omega/q$,
    \begin{equation}
        c_L = \sqrt{\dfrac{B+G}{\rho}}, \quad c_T = \sqrt{\dfrac{G}{\rho}}.
    \end{equation}
    These reduce to the standard longitudinal and transverse sound speeds when $k_{\phi\phi}=0$.

    \textbf{Poisson ratio.} To compute the Poisson ratio, we impose a uniaxial stress $\sigma_{yy}=p$ with $\sigma_{xx}=\sigma_{xy}=\sigma_{yx}=0$ and solve for the strain.
    We additionally set $u_{xy}=0$ to keep the top and bottom boundaries horizontal.
    Working in the basis $\sigma^a = C^{ab}u^b$, the constitutive matrix follows from Eq.~\eqref{eq:moduli_eff} with $A^o=K^o=0$ and (in the gauge of Eq.~\eqref{eq:dynmat_generic}) no coupling to rotations:
    \begin{equation}
        \begin{pmatrix} p \\ 0 \\ -p \\ 0 \end{pmatrix}
        =
        2\begin{pmatrix}
            B & 0 & 0 & 0 \\
            0 & 0 & 0 & 0 \\
            0 & 0 & G & 0 \\
            0 & 0 & 0 & G
        \end{pmatrix}
        \begin{pmatrix} u^0 \\ u^1 \\ u^2 \\ u^3 \end{pmatrix}.
    \end{equation}
    The last row gives $u^3=0$.
    Since $u^1=u_{yx}-u_{xy}$ and $u^3=u_{yx}+u_{xy}$, imposing $u_{xy}=0$ also fixes the rotation component to $u^1=0$.
    The remaining rows are diagonal, yielding
    \begin{equation}
        u^0 = \frac{p}{2B}, \quad u^2 = -\frac{p}{2G}, \quad u^1 = u^3 = 0.
    \end{equation}
    Recalling $u^a = u_{ij}\tau^a_{ij}$, we obtain
    \begin{equation}
        u_{xx} = \frac{p}{2}\frac{G-B}{BG}, \quad u_{yy} = \frac{p}{2}\frac{G+B}{BG}.
    \end{equation}
    Therefore,
    \begin{equation}
        \nu = -\frac{u_{xx}}{u_{yy}} = \frac{B-G}{B+G}.
    \end{equation}

    \subsection{The dispersion relation fitted to the measured spectra}
    \label{sec:dispersion_fit}

\begin{figure}[t]
            \centering
            \includegraphics{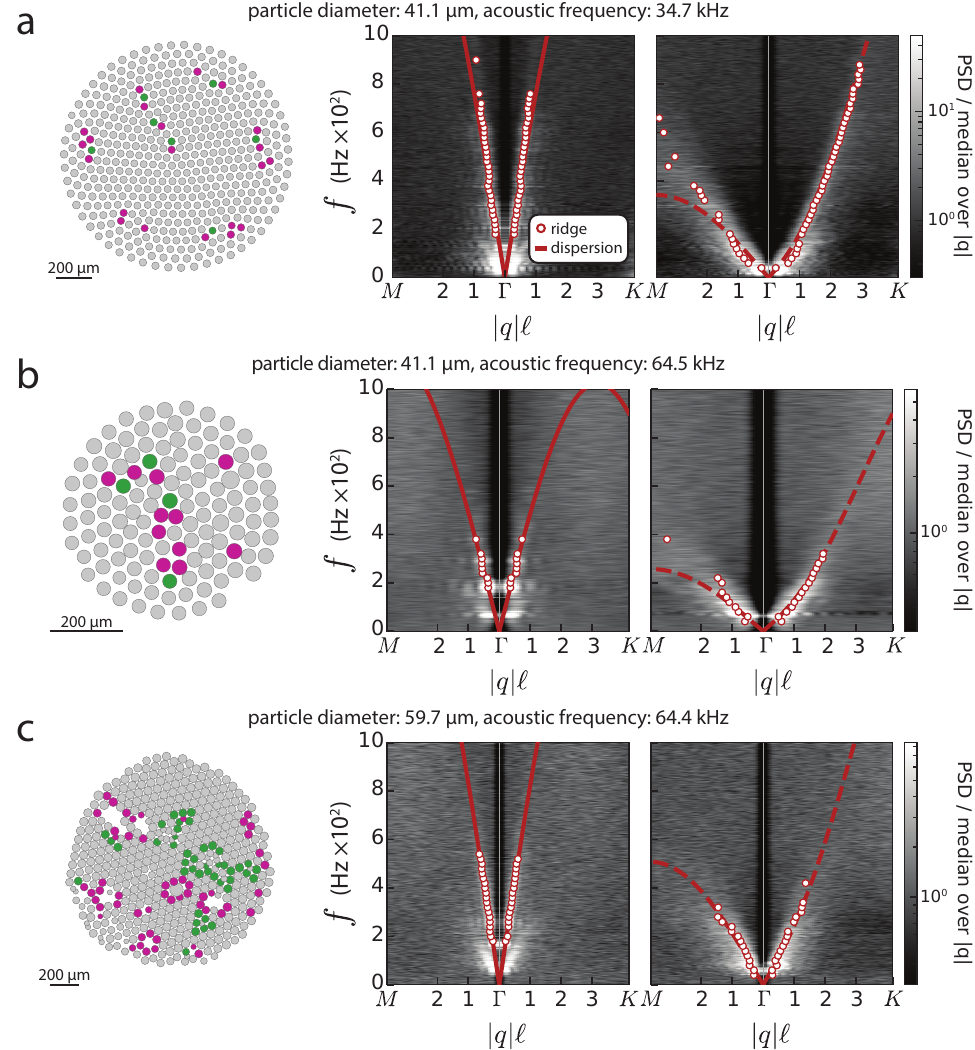}
            \caption{\textbf{Measured spectra and the fitted dispersion for the three raft-forming conditions.}
            Rows \textbf{a.}--\textbf{c.} correspond to conditions A, B, and D in Table~\ref{tab:dispersion_fits} (the three crystals in Fig.~4c of the main text).
            Left: one frame of the raft, with each particle drawn at the radius reported by the tracker.
            Interior sites with five neighbors are pink and those with seven are green.
            Scale bars are $200\,\mu\mathrm{m}$.
            Middle and right: longitudinal and transverse velocity-flux power spectra along $M \leftarrow \Gamma \rightarrow K$, computed in the co-rotating lattice frame so that wavevector directions remain fixed to the crystal axes.
            Each frequency row is normalized by its median over $|\bq|$, removing a wavevector-independent noise floor that rises with frequency.
            Red curves show the fitted dispersion of Eq.~\eqref{eq:dynmat_fit} (longitudinal: solid; transverse: dashed) using the parameters in Table~\ref{tab:dispersion_fits}.
            Circles mark the ridge points used in the fit.
            Frequencies assume a frame rate of $6000$~fps.
            }
            \label{fig_si:raft_spectra}
        \end{figure}

        The ridges of the measured power spectra are fitted to the dispersion that Eq.~\eqref{eq:dynmat_q} produces on the triangular lattice (Fig.~\ref{fig_si:raft_spectra}).
        With the chiral sector null and every stiffness measured in units of $k_{rr}$, that matrix depends on the single ratio $\kappa = k_{\phi \phi}/k_{rr}$,
        \begin{equation}
            \label{eq:dynmat_fit}
            D(\bq) = 2 k_{rr} \sum_{n=1}^{3}
            \left[ \ulattvec{n} \ulattvec{n} + \kappa \left( \mathsf{I} - \ulattvec{n} \ulattvec{n} \right) \right]
            \left[ 1 - \cos \left( \bq \cdot \lattvec{n} \right) \right],
        \end{equation}
        where $\mathsf{I} - \ulattvec{n} \ulattvec{n} = \ulattvecperp{n} \ulattvecperp{n}$ in two dimensions.

        We take the two branches by projecting $D(\bq)$ onto the propagation direction and onto its perpendicular,
        \begin{equation}
            \label{eq:branch_proj}
            \lambda_L = \hat{q}_i D_{ij} \, \hat{q}_j,
            \qquad
            \lambda_T = \hat{q}^\perp_i D_{ij} \hat{q}^\perp_j .
        \end{equation}
        Defining the natural angular frequency $\omega_0 = \sqrt{k_{rr}/m} \equiv 2\pi f_0$ with $m$ the particle mass, a plane wave of wavevector $\bq$ then has frequency
        \begin{equation}
            \label{eq:fit_law}
            f = f_0 \sqrt{\lambda(q\ell; \kappa)}.
        \end{equation}
        The frequency scale $f_0$ carries the only dimensional information, so each raft contributes two free parameters.
        They are fitted against four measured branches, the longitudinal and transverse ridges along $\Gamma$--K and along $\Gamma$--M.
        The $\Gamma$--K direction sits at $\theta = 0$ with zone edge $q\ell = 4\pi/3$, and $\Gamma$--M at $\theta = 30^\circ$ with zone edge $q\ell = 2\pi/\sqrt{3}$.

        The speed scale quoted in the main text follows from $\omega_0$ alone, $c_0 = \ell \omega_0$.
        Expanding Eq.~\eqref{eq:dynmat_fit} at small $q$ gives $\lambda_L \to 3(3+\kappa) (q\ell)^2 / 8$ and $\lambda_T \to 3(1+3\kappa) (q\ell)^2 / 8$, so that
        \begin{equation}
            \label{eq:c_from_fit}
            c_L = c_0 \sqrt{\frac{3(3+\kappa)}{8}},
            \qquad
            c_T = c_0 \sqrt{\frac{3(1+3\kappa)}{8}},
            \qquad
            \frac{c_T}{c_L} = \sqrt{\frac{1+3\kappa}{3+\kappa}} .
        \end{equation}
        The sound speeds are read off the $\vert \bq \vert \to 0$ limit of the fitted relation, with no separate fit to a linear region.
        Eq.~\eqref{eq:c_from_fit} agrees with $B$ and $G$ of Eq.~\eqref{eq:moduli_eff} through $c_L = \sqrt{(B+G)/\rho}$ and $c_T = \sqrt{G/\rho}$.
        The ratio $c_T / c_L$ depends on $\kappa$ alone, so the Poisson ratio is fixed by the fit without any knowledge of $k_{rr}$ or of the particle mass.
        The transverse branch softens as $\kappa \to -1/3$, which is the shear instability at $G = 0$ and bounds the grid search from below.

        A ridge point enters the fit under four conditions.
        It must not be pinned at the zone edge.
        The first-half and second-half averages of the Welch segments must agree to within $0.15$ in $q\ell$.
        Its contrast against the median over $\vert \bq \vert$ must exceed that of a position-permutation null by a factor of $1.3$.
        Longitudinal points are kept only above $170$~Hz, below which the longitudinal peak falls inside the finite-size aperture.
        The fit itself is a grid search over $\kappa \in (-1/3,\, 0.3]$ with $f_0$ solved analytically.

\begin{figure}[h]
            \centering
            \includegraphics{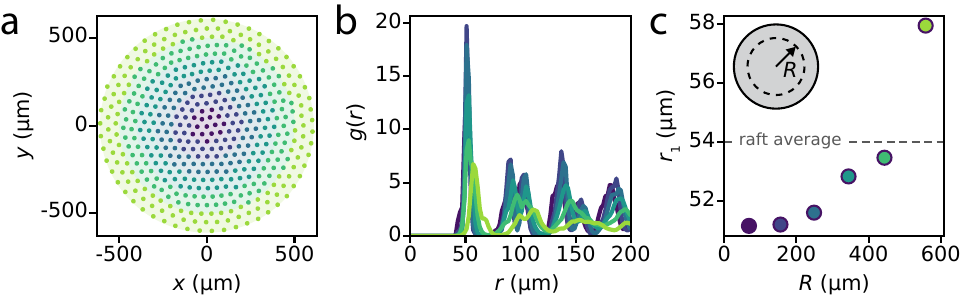}
            \caption{The lattice spacing of raft A grows toward its edge.
            \textbf{a.} Mean particle positions, colored by 100~$\mu$m ring from the center (dark) to the edge (light).
            \textbf{b.}The radial distribution function $g(r)$ of each ring, averaged over 1000 frames.
            \textbf{c.} The distance of the first peak of each $g(r)$, $r_1$, against the mean distance from the center in that bin.
            The dashed line is the raft-average spacing, the median Delaunay bond length, $54.0~\mu$m.}
            \label{fig_si:rdf}
        \end{figure}

        The raft is not uniformly spaced (Fig.~\ref{fig_si:rdf}).
        In raft A the nearest-neighbor spacing is $51.2~\mu$m within $200~\mu$m of the center and grows to $57.9~\mu$m in the outermost ring.
        We use the raft average throughout, defined as the median Delaunay bond length.
        For raft A we take it from the time-averaged positions, which gives $54.0~\mu$m.
        For the other rafts we take it from 20 frames spread over the record; applied to raft A, this procedure gives $54.1~\mu$m.
        Detections closer than half the median nearest-neighbor distance are first merged, since the tracker sometimes resolves one particle as two.
        This matters most for raft D, with about 20 merged pairs per frame.

        \begin{table}[h]
            \centering
            \caption{Dispersion fits for the five conditions.
            $d$ is the particle diameter and $f_d$ the drive frequency.
            $\ell$ is the raft-average spacing (Fig.~\ref{fig_si:rdf}).
            $N$ counts the ridge points entering the fit, and $\sigma_\kappa$ is the uncertainty from a moving-block bootstrap over them.
            Conditions C and E are excluded from the elastic analysis because the raft rearranges during the record.
            Conditions in which the particles did not assemble into a raft, such as the $25\,\mu\mathrm{m}$ particles of Fig.~4c of the main text, have no dispersion to fit and are not listed.}
            \label{tab:dispersion_fits}
            \begin{tabular}{l c c c r c r r c r c}
                \hline\hline
                Raft & $d$ ($\mu\mathrm{m}$) & $f_d$ ($\mathrm{kHz}$) & $\ell$ ($\mu\mathrm{m}$) & $\kappa$ & $s$ ($\mathrm{Hz}$) & $c_L$ ($\mathrm{mm/s}$) & $c_T$ ($\mathrm{mm/s}$) & $\nu$ & $N$ & $\sigma_\kappa$ \\
                \hline
                A & 41.1 & 34.7 & 54.0 & $-0.302$ & 788 & 269 & 50 & 0.930 & 120 & 0.0045 \\
                B & 41.1 & 64.5 & 50.7 & $-0.290$ & 504 & 162 & 35 & 0.904 & 43 & 0.0064 \\
                D & 59.7 & 64.4 & 63.4 & $-0.274$ & 852 & 343 & 88 & 0.869 & 66 & 0.0048 \\
                \hline
                C & 59.7 & 34.7 & 77.0 & $-0.285$ & 808 & 395 & 91 & 0.893 & 129 & 0.0056 \\
                E & 32.0 & 34.8 & 65.5 & $-0.302$ & 629 & 261 & 49 & 0.930 & 65 & 0.0055 \\
                \hline\hline
            \end{tabular}
        \end{table}

        The five values of $\kappa$ have mean $-0.291$ and standard deviation $0.012$, and the three raft-forming conditions alone give $-0.289$.
        All five sit well above the shear instability at $\kappa = -1/3$.
        The agreement with the trimer determination of Eq.~(4) of the main text is the central consistency check of the elastic analysis.
        These five conditions are the five rafts with measured spectra.
        Fig.~4c of the main text also shows five conditions, but a different set.
        It shows rafts A, B and D together with two conditions in which the particles never assemble, and it omits C and E.

        The transverse branch depends on direction in rafts A, B, C and E.
        Measured frequencies sit above the fit along $\Gamma$--M and below it along $\Gamma$--K.
        In raft A the mean residual in $\ln f$ is $+0.096$ along $\Gamma$--M and $-0.080$ along $\Gamma$--K, and the fit splits the difference.
        The split comes almost entirely from large wavevectors.
        In raft A it is $-0.003$ below $q\ell = 1.5$ and $+0.35$ above it, and rafts C and E give $+0.44$ and $+0.51$ above $q\ell = 1.5$.
        Raft D shows no split, but its transverse ridge ends near $q\ell = 1.5$, so it has no points where the split appears.
        A second neighbor shell is the natural suspect, since it changes the direction dependence of the branches at large $q$.
        With every second-shell coefficient taken from the three-particle simulations, it gives a split of the right sign that grows with $q$.
        The split it gives is a third to a half of the measured one, however.
        We therefore leave the remainder of the anisotropy unexplained.

    \subsection{The horizontal lines in the spectrum}
    \label{sec:stripes}

\begin{figure}[h]
            \centering
            \includegraphics[width=0.8\linewidth]{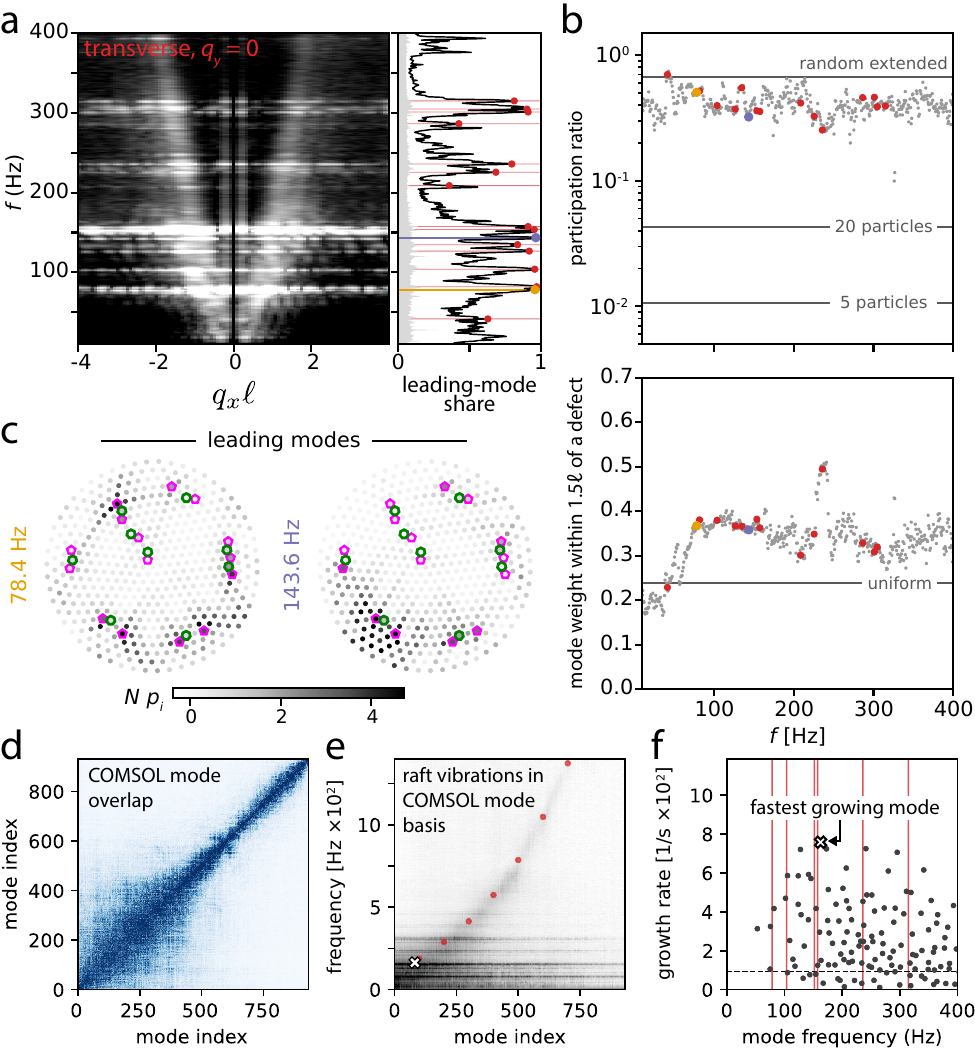}
            \caption{
            \textbf{Non-dispersive lines in the power spectrum are raft-wide collective oscillations.}
            \textbf{a.} Left: transverse spectrum of raft A along $q_y = 0$ as a function of wavevector $q_x \ell$ ($\ell$ is the lattice spacing) and frequency $f$. Horizontal bright lines are oscillations that appear at a fixed frequency for all wavevectors. Right: leading-mode share at each frequency, defined as the fraction of the raft's power at that frequency carried by its single strongest spatial pattern (the leading singular vector of the particle velocities across time windows). Red points and lines mark the frequencies of the spectral lines; the orange and purple points mark the 78.4 Hz and 143.6 Hz lines shown in (c). At the lines, nearly all of the motion is in one pattern.
            \textbf{b.} Top: participation ratio of the leading mode at each frequency (gray; colored points as in (a)).
            Horizontal lines show the values for a random pattern spread over the whole raft and for patterns confined to 20 or 5 particles.
            Bottom: fraction of the leading mode's weight on particles within $1.5\ell$ of a topological defect.
            The horizontal line is the value for a mode spread uniformly over all particles. 
            \textbf{c.} Leading modes at 78.4 Hz and 143.6 Hz.
            Each particle is shaded by $N p_i$, where $p_i$ is its share of the mode's weight and $N$ is the number of particles ($N p_i = 1$ for uniform motion).
            Pink and green symbols mark five- and seven-fold coordinated particles.
            \textbf{d.} Overlap $|\hat{e}_i \cdot \hat{e}_j|$ between the eigenmodes of the dynamical matrix obtained from finite-element (COMSOL) simulations, ordered by eigenfrequency.
            \textbf{e.} Power spectrum of the measured particle motion projected onto each COMSOL eigenmode.
            Red points show the eigenfrequencies of every hundredth mode.
            The white cross marks the mode with the largest growth rate.
            The measured motion follows the predicted eigenfrequencies at high frequency, while the spectral lines appear as horizontal bands across all modes.
            \textbf{f.} Growth rate of the COMSOL eigenmodes against their frequency, in the absence of damping. Red lines mark the measured line frequencies, the dashed line is the growth rate required to overcome air drag, and the cross marks the fastest-growing mode.
            }
            \label{fig_si:stripe_modes}
        \end{figure}

        The measured spectra contain horizontal lines, frequencies at which power appears at every wavevector (Fig.~\ref{fig_si:stripe_modes}).
        We show here that these are collective modes, rather than the result of individual particle motions.
        
        First, we form the cross-spectral matrix of all particle velocities at each frequency from 58 Hann-windowed segments of 4096 frames.
        We call the eigenvector associated with the largest eigenvalue, $\lambda_{\max}$, the leading-mode.
        To measure how much that single eigenvector contributes to the motion of the raft at that frequency, we define the mode share of mode $n$ as $\lambda_n / \sum_m \lambda_m$.
        At 13 of the 16 detected lines, the leading-mode carries between $69\%$ and $97\%$ of the motion.
        The same calculation on phase-randomized data keeps every particle's spectrum but removes the coherence between particles, giving a share of $7\%$ at the median frequency and $12\%$ at the 95th percentile (SI Fig.~\ref{fig_si:raft_spectra}a).

        We also count the particles that take part in each mode with the participation ratio $\mathrm{PR} = \left(\sum_i \vert \mathbf{u}_i \vert^2\right)^2 / \left(N \sum_i \vert \mathbf{u}_i \vert^4\right)$, where $\mathbf{u}_i$ is the two-component $(x, y)$ amplitude of particle $i$ in the leading mode and $N = 466$.
        A mode on $m$ particles has $\mathrm{PR} \approx m/N$, and a random extended pattern has $\mathrm{PR} \approx 0.67$.
        At the 13 lines $\mathrm{PR} = 0.25$ to $0.55$, so each mode extends over a quarter to half of the raft.
        Frequencies between the lines give similar values, so no frequency in this range carries a localized mode.
        The modes lean mildly toward defects.
        Particles within $1.5\ell$ of an interior 5- or 7-fold site make up $24\%$ of the raft and carry $36\%$ of a typical line mode, against $34\%$ between the lines (SI Fig.~\ref{fig_si:stripe_modes}c, b).

        We hypothesize that these lines are internally driven by nonreciprocal forces acting on non-normal eigenmodes.
        Using COMSOL simulations of the entire raft, we measure significant non-normality of the dynamical matrix (SI Fig.~\ref{fig_si:stripe_modes}d).
        Projecting the raft motion onto these modes, we see that the resulting spectrum has a dispersion relation that matches the COMSOL mode frequencies (SI Fig.~\ref{fig_si:stripe_modes}e).
        Additionally, due to the non-reciprocity of the system, the eigenfrequencies will be complex, which denote the growth/decay rate of the modes. 
        The dominant spectral line, at 170 Hz, coincides with the frequency of the fastest-growing mode in the undamped COMSOL calculation (Supplementary Fig.~\ref{fig_si:stripe_modes}f).
        Meanwhile, due to the non-orthogonality of the system, the dominant mode `leaks' into all k by mode-coupling, producing a horizontal line at the predicted frequency. 
        Nonlinear interactions may generate some of the other spectral lines at subharmonics or higher harmonics of this mode.

\section{Nonreciprocity is visible at broken six-fold symmetry}
\label{sec:sources}

    Writing the dynamical matrix as $D^{\alpha\beta}_{ij} = -\partial F^\alpha_i / \partial u^\beta_j$, we have $D = D^+ + D^-$ with $D^\pm = (D \pm D^{\mathsf T})/2$ and $(D^{\mathsf T})_{ij}^{\alpha \beta} = D_{ji}^{\beta \alpha}$.
    The antisymmetric part $D^-$ is the failure of Maxwell--Betti reciprocity, and $\Vert D^-_i \Vert^2 = \sum_{j, \alpha, \beta} [(D^-)_{ij}^{\alpha \beta}]^2$ denotes the Frobenius norm (squared) of particle $i$'s row of $2\times2$ blocks.

    \subsection{The nonreciprocity is proportional to \texorpdfstring{$\alpha_3$}{alpha3} alone}
    \label{sec:A_alpha3}

        Ignoring the chiral sector of the force law, the antisymmetric part of the dynamical matrix can be seen to depend solely on $\alpha_3$ based on two observations.
        First, the force law is linear in the coefficients $(a_0, k_a, a_3, \alpha_3)$, so $D^-$ is linear in them.
        Second, setting $\alpha_3 = 0$ leaves $a_{ijk} = a_0 + k_a \delta r_{ij} + a_3 (\delta r_{ik} + \delta r_{jk})$, for which $f_2 = a_3$ is constant and the integrability condition Eq.~\eqref{eq:integrability} is satisfied identically.
        The force is then conservative, its Jacobian is symmetric, and $D^- = 0$ in all configurations.
        Together these give
        \begin{equation}
            \label{eq:A_is_alpha3}
            D^- = \alpha_3 \, D^{-(\alpha_3)}(\{\br\}),
        \end{equation}
        with $D^{-(\alpha_3)}$, the antisymmetric part evaluated at unit $\alpha_3$, encoding the configuration's geometry.

    \subsection{Prediction of nonreciprocity with strain amplitude}
    \label{sec:selection_rule}
        We write the local deformation gradient in the orthonormal basis
        \begin{equation}
            \label{eq:generators}
            \nabla \mathbf{u} = \sum_\alpha c_\alpha \tau^\alpha,
            \qquad
            \tau^{\rm dil} = \begin{pmatrix} 1 & 0 \\ 0 & 1\end{pmatrix}, \;
            \tau^{\rm rot} = \begin{pmatrix} 0 & -1 \\ 1 & 0\end{pmatrix}, \;
            \tau^{1} = \begin{pmatrix} 1 & 0 \\ 0 & -1\end{pmatrix}, \;
            \tau^{2} = \begin{pmatrix} 0 & 1 \\ 1 & 0\end{pmatrix},
        \end{equation}
        which are orthonormal under the inner product $(u, v) = \mathrm{Tr}(u^{\mathsf T} v)/2$, so that $c_\alpha = \tfrac{1}{2}\mathrm{Tr}(\nabla \mathbf{u}^{\mathsf T} \tau^\alpha)$.
        Writing the displacement gradient components as $u_{ij} = \partial u_i / \partial x_j$, we have $c_{\rm dil} = (u_{xx} + u_{yy})/2$, $c_{\rm rot} = (u_{yx} - u_{xy})/2$, $c_1 = (u_{xx} - u_{yy})/2$ and $c_2 = (u_{xy} + u_{yx})/2$.
        Under the hexagonal point group of the lattice $C_{6v}$, these four are not interchangeable.
        The dilation is the invariant $A_1$, the rotation the pseudoscalar $A_2$, and the two shears form the doublet $E_2$~\cite{Born1954}.

        The nonreciprocal three-body force $\alpha_3$ couples only to $E_2$.
        In words: a uniform dilation or a rigid rotation of the lattice never produces nonreciprocity, while a shear does, by an amount proportional to the size of the shear and independent of its direction.
        We show this by computing $D^-$ for an interior particle $i$ of a uniformly deformed lattice, to first order in $\nabla \mathbf{u}$.
        The calculation is short once two pieces of bookkeeping are in place: the geometry of the six-fold neighbor shell, and the way a uniform deformation moves a single bond.

        Let $\hat{\mathbf{r}}_n = \left( \cos\theta_n, \sin\theta_n \right)$, $n = 1, \dots, 6$, be the unit vectors from particle $i$ to its six neighbors, with $\theta_n = (n-1)2\pi/6$.
        The outer product of a bond with itself splits over the basis of Eq.~\eqref{eq:generators} as
        \begin{equation}
            \label{eq:bond_outer}
            \hat{\mathbf{r}}_n\hat{\mathbf{r}}_n^{\mathsf T} = \begin{pmatrix}
                \cos^2\theta_n & \cos\theta_n \sin\theta_n \\
                \cos\theta_n \sin\theta_n & \sin^2\theta_n
            \end{pmatrix}
            =
            \dfrac{1}{2} \begin{pmatrix}
                1+\cos 2 \theta_n & \sin 2\theta_n \\
                \sin 2\theta_n & 1 - \cos 2\theta_n
            \end{pmatrix}
            =
            \dfrac{1}{2} \left( \tau^\mathrm{dil} + \cos 2\theta_n \, \tau^1 + \sin 2\theta_n \, \tau^2\right).
        \end{equation}
        The doubled angle $2 \theta_n$ indicates that a bond and its reverse give the same matrix.
        Sums over the shell therefore produce sums of the form $\sum_n e^{i m \theta_n}$ with $m$ even, and for a six-fold shell
        \begin{equation}
            \label{eq:shell_sums}
            \sum_{n=1}^{6} e^{2 i \theta_n} = 0,
            \qquad
            \sum_{n=1}^{6} e^{4 i \theta_n} = 0,
            \qquad
            \sum_{n=1}^{6} e^{6 i \theta_n} = 6.
        \end{equation}
        Averaging Eq.~\eqref{eq:bond_outer} over the shell and using the first sum gives $\langle \hat{\mathbf{r}}_n\hat{\mathbf{r}}_n^{\mathsf T} \rangle = \mathsf{I}/2$.
        In other words, the triangular lattice is isotropic.

        Under a uniform deformation each bond vector changes as $\mathbf{r}_n \to (\mathsf{I} + \nabla \mathbf{u}) \, \mathbf{r}_n$.
        To first order, the direction of the bond turns by the angle
        \begin{equation}
            \label{eq:bond_turn}
            \delta\theta_n = \hat{\mathbf{r}}_n^{\perp} \left( \nabla \mathbf{u} \, \hat{\mathbf{r}}_n \right) = c_{\rm rot} + c_2 \cos 2\theta_n - c_1 \sin 2\theta_n,
        \end{equation}
        where the second equality follows from evaluating $\hat{\mathbf{r}}_n^{\perp} \cdot ( \tau^\alpha \hat{\mathbf{r}}_n )$ for each generator, which gives $0$, $1$, $-\sin 2\theta_n$ and $\cos 2\theta_n$ for $\alpha = {\rm dil}, {\rm rot}, 1, 2$.
        The three kinds of deformation act differently.
        A dilation changes the length of every bond but turns none of them.
        A rotation turns every bond by the same angle $c_{\rm rot}$.
        Shears turn different bonds by different amounts, depending on the shear direction.
        We combine the two shear components into a single complex number, $c_1 + i c_2 = \vert \varepsilon_{\rm dev} \vert \, e^{2 i \phi}$, where $\vert \varepsilon_{\rm dev} \vert$ is the size of the shear and $\phi$ is the direction along which it stretches.
        A shear axis is defined only modulo $180^\circ$, so, as for a bond, the phase is twice the physical angle.
        In this notation Eq.~\eqref{eq:bond_turn} reads
        \begin{equation}
            \label{eq:bond_turn_phase}
            \delta\theta_n = c_{\rm rot} + \vert \varepsilon_{\rm dev} \vert \sin 2(\phi - \theta_n).
        \end{equation}

        \textbf{Step 1: each nonreciprocal block is a pure twist.}
        Consider the block of $D$ that couples $i$ to a neighbor $j$.
        A rotation by $\pi$ about the midpoint of the bond $ij$ maps a uniformly deformed lattice onto itself and exchanges $i$ and $j$.
        The force law is covariant under rotations, and a rotation by $\pi$ reverses both the forces and the displacements, so it leaves every block of the dynamical matrix unchanged.
        Hence $D_{ij} = D_{ji}$, and
        \begin{equation}
            \label{eq:twist_def}
            (D^-)_{ij} = \left( D_{ij} - D_{ji}^{\mathsf T} \right)/2 = \left( D_{ij} - D_{ij}^{\mathsf T} \right)/2 = w_{ij} \, \tau^{\rm rot},
            \qquad
            w_{ij} = \mathrm{Tr}\left( D_{ij}^{\mathsf T} \tau^{\rm rot} \right)/2.
        \end{equation}
        The first equality is the definition of $D^-$, the second uses $D_{ji} = D_{ij}$, and the third uses the fact that every antisymmetric $2 \times 2$ matrix is a multiple of $\tau^{\rm rot}$.
        The nonreciprocity of a bond is therefore a single number, $w_{ij}$, which is the $\tau^{\rm rot}$ component of the block in the basis of Eq.~\eqref{eq:generators}.
        We call it the twist of the bond.
        At a free edge the rotation about the bond midpoint is not a symmetry and this step fails (we return to this at the end of the section).

        By Eq.~\eqref{eq:A_is_alpha3}, only $\alpha_3$ contributes to $w_{ij}$.
        In a triangle with vertices $i$, $j$ and $k$, the $\alpha_3$ part of the force on $i$ is Eq.~\eqref{eq:coeffs} summed over the two ordered pairs $(j, k)$ and $(k, j)$,
        \begin{equation}
            \mathbf{F}_i^{\alpha_3} = \alpha_3 \left[ \left( \delta r_{ik} - \delta r_{jk} \right) \hat{\mathbf{r}}_{ij} + \left( \delta r_{ij} - \delta r_{jk} \right) \hat{\mathbf{r}}_{ik} \right].
        \end{equation}
        We differentiate with respect to $\mathbf{r}_j$, using $\partial \delta r_{ij} / \partial \mathbf{r}_j = \hat{\mathbf{r}}_{ij}$, $\partial \delta r_{jk} / \partial \mathbf{r}_j = -\hat{\mathbf{r}}_{jk}$ and $\partial \delta r_{ik} / \partial \mathbf{r}_j = 0$, which gives
        \begin{equation}
            \dfrac{\partial \mathbf{F}_i^{\alpha_3}}{\partial \mathbf{r}_j} = \alpha_3 \left[ \hat{\mathbf{r}}_{ij} \hat{\mathbf{r}}_{jk}^{\mathsf T} + \hat{\mathbf{r}}_{ik} \left( \hat{\mathbf{r}}_{ij} + \hat{\mathbf{r}}_{jk} \right)^{\mathsf T} \right]
            + \alpha_3 \left( \delta r_{ik} - \delta r_{jk} \right) \dfrac{\mathsf{I} - \hat{\mathbf{r}}_{ij} \hat{\mathbf{r}}_{ij}^{\mathsf T}}{r_{ij}}.
        \end{equation}
        The last term comes from differentiating the direction $\hat{\mathbf{r}}_{ij}$; it is symmetric and so carries no twist.
        For the remaining terms we need the $\tau^{\rm rot}$ component of an outer product $\hat{\mathbf{a}} \hat{\mathbf{b}}^{\mathsf T}$ of two unit vectors at angles $\theta_a$ and $\theta_b$, which is $\tfrac{1}{2} \mathrm{Tr}\big( (\hat{\mathbf{a}} \hat{\mathbf{b}}^{\mathsf T})^{\mathsf T} \tau^{\rm rot} \big) = \tfrac{1}{2} \, \hat{\mathbf{a}} \cdot \hat{\mathbf{b}}^\perp = -\tfrac{1}{2} \sin(\theta_b - \theta_a)$.
        Using $D_{ij} = -\partial \mathbf{F}_i / \partial \mathbf{r}_j$ and adding the two triangles $T$ and $T'$ that share the bond $ij$,
        \begin{equation}
            \label{eq:twist_angles}
            w_{ij} = \dfrac{\alpha_3}{2} \sum_{k \in \{T, T'\}} \left[ \sin(\theta_{jk} - \theta_{ij}) + \sin(\theta_{ij} - \theta_{ik}) + \sin(\theta_{jk} - \theta_{ik}) \right],
        \end{equation}
        where $\theta_{ab}$ is the direction of the bond from $a$ to $b$.
        Eq.~\eqref{eq:twist_angles} contains only differences of directions between bonds.
        This is the selection rule.
        A dilation changes no direction, so it cannot produce a twist.
        A rotation changes every direction by the same amount, which leaves every difference unchanged, so it cannot produce a twist either.
        Both statements hold exactly, not only to first order.
        Only a shear, which turns different bonds by different amounts, can make $w_{ij}$ nonzero.

        On the undeformed lattice, take $\theta_{ij} = \theta$.
        In the triangle whose third vertex $k$ lies counterclockwise of the bond, $\theta_{ik} = \theta + 60^\circ$ and $\theta_{jk} = \theta + 120^\circ$, and the bracket in Eq.~\eqref{eq:twist_angles} is $\sin 120^\circ - \sin 60^\circ + \sin 60^\circ = \sqrt{3}/2$.
        The other triangle is the mirror image of the first across the bond, with $\theta_{ik'} = \theta - 60^\circ$ and $\theta_{jk'} = \theta - 120^\circ$, and its bracket is $-\sqrt{3}/2$.
        The two cancel, so $w_{ij} = 0$ and the undeformed lattice is reciprocal, $D^{-(\alpha_3)} = 0$.
        This cancellation is the mirror symmetry of the lattice at work.

        Under a deformation, we expand each sine to first order about its undeformed argument.
        In each bracket the changes in $\theta_{jk}$ cancel, and what remains is $\delta\theta_{ij} - \delta\theta_{ik}$, so
        \begin{equation}
            w_{ij} = \dfrac{\alpha_3}{2} \left( 2 \, \delta\theta_{ij} - \delta\theta_{ik} - \delta\theta_{ik'} \right),
        \end{equation}
        where $ik$ and $ik'$ are the two bonds of the shell adjacent to $ij$, at $\theta \pm 60^\circ$.
        Substituting Eq.~\eqref{eq:bond_turn_phase}, the rotation $c_{\rm rot}$ cancels because $2 - 1 - 1 = 0$.
        The shear terms combine as $2 \sin x - \sin(x - 120^\circ) - \sin(x + 120^\circ) = 3 \sin x$, with $x = 2(\phi - \theta)$.
        The result is
        \begin{equation}
            \label{eq:twist}
            w_{ij} = \dfrac{3}{2} \, \alpha_3 \, \vert \varepsilon_{\rm dev} \vert \sin 2(\phi - \theta_{ij})
            = \dfrac{3}{2} \, \alpha_3 \left( \delta\theta_{ij} - c_{\rm rot} \right).
        \end{equation}
        The twist on a bond is proportional to the angle through which the shear turns that bond.
        Bonds along the principal axes of the shear carry no nonreciprocity, and bonds at $45^\circ$ to them carry the most.

        Translation invariance gives $D_{ii} = -\sum_j D_{ij}$, so $(D^-)_{ii} = -\left( \sum_n w_n \right) \tau^{\rm rot}$, where $w_n$ is the twist of bond $n$.
        Writing Eq.~\eqref{eq:twist} as $w_n = \tfrac{3}{2} \alpha_3 \, \mathrm{Im}\big[ (c_1 + i c_2) \, e^{-2 i \theta_n} \big]$, the sum is proportional to $\sum_n e^{-2 i \theta_n}$, which vanishes by Eq.~\eqref{eq:shell_sums}.
        The self block is therefore reciprocal, and all of the nonreciprocity in particle $i$'s row sits in the six bond twists.

        \textbf{Step 5: the size of the nonreciprocity.}
        Each bond block contributes $\Vert w_n \tau^{\rm rot} \Vert^2 = 2 w_n^2$ to the squared Frobenius norm, so
        \begin{equation}
            \Vert D^-_i \Vert^2 = 2 \sum_n w_n^2 = \dfrac{9}{2} \, \alpha_3^2 \, \vert \varepsilon_{\rm dev} \vert^2 \sum_n \sin^2 2(\phi - \theta_n).
        \end{equation}
        Writing $\sin^2 x = (1 - \cos 2x)/2$, the sum becomes $3 - \tfrac{1}{2} \mathrm{Re} \big[ e^{4 i \phi} \sum_n e^{-4 i \theta_n} \big] = 3$, by the second of Eqs.~\eqref{eq:shell_sums}.
        Therefore
        \begin{equation}
            \label{eq:A_vs_strain}
            \Vert D^-_i \Vert = \frac{3\sqrt{6}}{2} \, \vert \alpha_3 \vert \, \vert \varepsilon_{\rm dev} \vert + \mathcal{O}(\varepsilon^2),
            \qquad
            \vert \varepsilon_{\rm dev} \vert = \sqrt{c_1^2 + c_2^2}.
        \end{equation}
        Each individual twist depends on the shear direction $\phi$, but the sum of their squares does not, because the second shell sum vanishes.
        In the language of the point group, the only quantity quadratic in the $E_2$ doublet that is invariant under $C_{6v}$ is $c_1^2 + c_2^2$.
        A dependence on $\phi$ would first come through the third shell sum, which needs sixth harmonics of the bond angles.
        The first-order twists contain only second harmonics, so a $\cos 6\phi$ anisotropy can enter only at higher order in the strain.
        Because the twist depends only on bond directions, the constant $3\sqrt{6}/2$ is a pure number: it does not depend on the lattice spacing, on the reference length of the force law, or on which interior particle is chosen.

        The chiral terms of the force law point along $\hat{\mathbf{r}}^\perp_{ij}$, which changes sign under a reflection.
        With any of $b_0$, $k_b$, $b_3$ or $\beta_3$ switched on, the two triangles flanking a bond no longer cancel in Step 3, and $D^- \neq 0$ already on the undeformed lattice.
        This is why the selection rule is stated for the achiral model.

        The abscissa of Fig.~5b measures the same quantity directly from particle positions, without reference to a displacement field.
        Define the shell tensor $\mathsf{S}_i = \langle \mathbf{d}_{ij} \otimes \mathbf{d}_{ij}\rangle / \ell^2$, where $\mathbf{d}_{ij}$ is the bond vector from $i$ to neighbor $j$, the average runs over the six neighbors, and $\ell$ is the lattice spacing.
        For the ideal shell, $\mathbf{d}_{ij} = \ell \, \hat{\mathbf{r}}_n$, and the shell average below Eq.~\eqref{eq:shell_sums} gives $\mathsf{S}_i = \mathsf{I}/2$ exactly.
        Under a uniform deformation $\mathbf{d}_{ij} = \ell \, (\mathsf{I} + \nabla \mathbf{u}) \hat{\mathbf{r}}_n$, and to first order
        \begin{equation}
            \label{eq:shell_tensor_strain}
            \mathsf{S}_i = \tfrac{1}{2} \mathsf{I} + \tfrac{1}{2} \left( \nabla \mathbf{u} + \nabla \mathbf{u}^{\mathsf T} \right)
            = \left( \tfrac{1}{2} + c_{\rm dil} \right) \tau^{\rm dil} + c_1 \tau^1 + c_2 \tau^2 .
        \end{equation}
        The rotation drops out because it is antisymmetric, the dilation enters only through the trace, and the traceless (deviatoric) part of $\mathsf{S}_i$ has components exactly $(c_1, c_2)$, with size $\vert \varepsilon_{\rm dev} \vert$.
        The abscissa of Fig.~5b therefore measures how far the local deformation has left the $A_1 \oplus A_2$ subspace.

        The three predictions --- no response to dilation or rotation, equal response to the two shears, and no dependence on the shear direction $\phi$ --- are checked separately at strain amplitude $\varepsilon = 10^{-2}$ on the interior of a perfect disk.
        The two kernel directions sit at the finite-difference floor, $\Vert D^- \Vert / \Vert D^+ \Vert = 2.3\times10^{-11}$ for dilation and $2.5\times10^{-11}$ for rotation, at every amplitude tested.
        The two shears agree with each other, $1.589\times10^{-5}$ and $1.593\times10^{-5}$.
        Carrying the doublet phase through a full period at fixed $\vert \varepsilon_{\rm dev} \vert$ leaves the response constant to $0.50\%$ peak to peak, the residual being dominated by a $\cos 6\phi$ modulation of relative amplitude $2.5\times10^{-3}$ --- the first anisotropy the six-fold shell admits (Step 5). It is a higher-order effect, as expected: in an ideal interior its relative amplitude is $\varepsilon/4$, growing linearly with the strain.
        The weaker $\cos 2\phi$, $\cos 4\phi$ and $\cos 8\phi$ terms, at or below $1.2\times10^{-3}$, are consistent with the finite disk and the circular interior mask, neither of which is six-fold symmetric.

        Eq.~\eqref{eq:A_vs_strain} is a statement about the \emph{numerator}.
        Fig.~5b of the main text plots the ratio $\Vert D^-_i \Vert / \Vert D^+_i \Vert$, and its slope inherits the achiral stiffnesses through the denominator.
        The denominator can also be written in closed form.
        We collect the four achiral coefficients into the vector $\mathbf{x} = (a_0/\ell,\, k_a,\, a_3,\, \alpha_3)$, whose entries all have units of force per length.
        We derive $\Vert D^+_i \Vert^2$ for an interior particle of the undeformed lattice in three steps.

        First, we write down the row of $D^+$ that belongs to particle $i$.
        Every triangle that contains $i$ is made of $i$ and two of its nearest neighbors, so the row has only seven nonzero $2 \times 2$ blocks.
        These are one block coupling $i$ to each of its six neighbors and one self block.
        Let $\hat{\mathbf{n}}_j$ be the unit vector from $i$ to neighbor $j$, and $\hat{\mathbf{p}}_j$ its counterclockwise rotation by $90^\circ$.
        In the bond frame the force law is Eq.~(3) of the main text, with $k_{r\phi} = k_{\phi r} = 0$ for the achiral model.
        Rotating back to the lab frame gives the block that couples $i$ to $j$,
        \begin{equation}
            \label{eq:bond_block}
            (D^+)_{ij} = -\left( k_{rr} \, \hat{\mathbf{n}}_j \hat{\mathbf{n}}_j^{\mathsf T} + k_{\phi\phi} \, \hat{\mathbf{p}}_j \hat{\mathbf{p}}_j^{\mathsf T} \right).
        \end{equation}
        Translation invariance makes the self block minus the sum of the six bond blocks.
        For six bonds $60^\circ$ apart, $\sum_j \hat{\mathbf{n}}_j \hat{\mathbf{n}}_j^{\mathsf T} = \sum_j \hat{\mathbf{p}}_j \hat{\mathbf{p}}_j^{\mathsf T} = 3 \mathsf{I}$, so
        \begin{equation}
            \label{eq:self_block}
            (D^+)_{ii} = 3 \left( k_{rr} + k_{\phi\phi} \right) \mathsf{I}.
        \end{equation}

        Second, we add up the squared entries of these seven blocks.
        A rotation does not change the Frobenius norm, so each bond block contributes $k_{rr}^2 + k_{\phi\phi}^2$, and the six together contribute $6 ( k_{rr}^2 + k_{\phi\phi}^2 )$.
        The self block has two equal diagonal entries and contributes $2 \times 9 ( k_{rr} + k_{\phi\phi} )^2$.
        The sum is
        \begin{equation}
            \label{eq:S_norm_K}
            \Vert D^+_i \Vert^2 = 24 k_{rr}^2 + 36 k_{rr} k_{\phi\phi} + 24 k_{\phi\phi}^2 .
        \end{equation}

        Third, we express $k_{rr}$ and $k_{\phi\phi}$ in terms of $\mathbf{x}$.
        Eq.~(4) of the main text writes each as a fixed linear combination of the coefficients,
        \begin{equation}
            \label{eq:Kmap_vectors}
            k_{rr} = \mathbf{r} \cdot \mathbf{x}, \quad \mathbf{r} = \left( 0,\, 2,\, \tfrac{5}{2},\, -\tfrac{1}{2} \right),
            \qquad
            k_{\phi\phi} = \mathbf{p} \cdot \mathbf{x}, \quad \mathbf{p} = \left( 2,\, 0,\, -\tfrac{3}{2},\, \tfrac{3}{2} \right).
        \end{equation}
        The coefficient $\alpha_3$ appears in both, because on the undeformed lattice its contribution to the force has a symmetric part, which renormalizes $k_{rr}$ and $k_{\phi\phi}$.
        Substituting Eq.~\eqref{eq:Kmap_vectors} into Eq.~\eqref{eq:S_norm_K} gives a quadratic form in $\mathbf{x}$,
        \begin{equation}
            \label{eq:S_quadratic}
            \Vert D^+_i \Vert^2 = \mathbf{x}^{\mathsf T} \mathsf{Q} \,\mathbf{x},
            \qquad
            \mathsf{Q} = 24 \, \mathbf{r} \mathbf{r}^{\mathsf T} + 18 \left( \mathbf{r} \mathbf{p}^{\mathsf T} + \mathbf{p} \mathbf{r}^{\mathsf T} \right) + 24 \, \mathbf{p} \mathbf{p}^{\mathsf T}
            =
            \begin{pmatrix}
                96 & 72 & 18 & 54 \\
                72 & 96 & 66 & 30 \\
                18 & 66 & 69 & -3 \\
                54 & 30 & -3 & 33
            \end{pmatrix}.
        \end{equation}
        For example, the top-left entry is $24 r_1^2 + 36 r_1 p_1 + 24 p_1^2 = 24 \times 2^2 = 96$, since $r_1 = 0$ and $p_1 = 2$.

        The same matrix has a second reading, which is how we check it numerically.
        The force law is linear in the coefficients, so $D^+$ is too, and $D^+ = \sum_\alpha x_\alpha D^{+(\alpha)}$, where $D^{+(\alpha)}$ is the symmetric part computed with only coefficient $\alpha$ switched on at unit value.
        Then $\Vert D^+_i \Vert^2 = \sum_{\alpha \beta} x_\alpha x_\beta \langle D^{+(\alpha)}_i, D^{+(\beta)}_i \rangle$, where $\langle \cdot, \cdot \rangle$ is the Frobenius inner product of two rows.
        So $\mathsf{Q}_{\alpha\beta} = \langle D^{+(\alpha)}_i, D^{+(\beta)}_i \rangle$ is the Gram matrix of the four single-coefficient rows.
        Computing each $D^{+(\alpha)}$ by finite differences of the implemented force law on a perfect disk and forming this Gram matrix reproduces the integers in Eq.~\eqref{eq:S_quadratic} to $3\times10^{-9}$.
        The quadratic form also matches the measured $\Vert D^+_i \Vert$ to a relative $3\times10^{-11}$ over independent variations of all four coefficients.

        Two properties follow from the derivation.
        $\mathsf{Q}$ is positive semidefinite, because Eq.~\eqref{eq:S_norm_K} is a positive definite form in $(k_{rr}, k_{\phi\phi})$.
        $\mathsf{Q}$ has rank two, because $\Vert D^+_i \Vert$ depends on the four coefficients only through the two combinations $k_{rr}$ and $k_{\phi\phi}$.
        Its two nonzero eigenvalues are $82.0$ and $212.0$.
        Dividing Eq.~\eqref{eq:A_vs_strain} by the square root of Eq.~\eqref{eq:S_quadratic} gives the slope plotted in Fig.~5b,
        \begin{equation}
            \label{eq:slope_closed}
            C = \frac{3\sqrt{6}}{2} \, \frac{\vert \alpha_3 \vert}{\sqrt{\mathbf{x}^{\mathsf T} \mathsf{Q} \mathbf{x}}} .
        \end{equation}
        Both numerator and denominator are homogeneous of degree one in the coefficients, so $C$ is scale free: it depends only on the dimensionless ratios $a_0/(\ell k_a)$, $a_3/k_a$ and $\alpha_3/k_a$ and can be evaluated straight from the measured table.
        Doing so gives $C = 1.593 \times 10^{-3}$ per unit deviatoric strain against $1.589 \times 10^{-3}$ measured at $\varepsilon = 10^{-2}$, the $0.3\%$ difference being the second-order term, which vanishes as $\varepsilon \to 0$ rather than indicating a discrepancy.
        The dashed line in Fig.~5b is therefore a parameter-free prediction, not a fit.
        The free edge of a finite raft is nonreciprocal at zero strain --- the whole disk gives $\Vert D^- \Vert / \Vert D^+ \Vert = 3.9\times10^{-4}$ with no strain applied at all --- so the strain law is a bulk statement and is extracted on the interior, or under periodic boundaries.

    \subsection{The force on a dislocation lies along climb}
    \label{sec:climb}

        As derived above, the net force over a triplet depends only on $\alpha_3$ and $\beta_3$.
        The direction is set by the mirror symmetry of those two terms.
        An edge dislocation with Burgers vector $\mathbf{b}$ is invariant under reflection about its climb axis $\hat{\mathbf{z}} \times \mathbf{b}$.
        With $\mathbf{b}$ along $x$, the strain components $u_{xx}$ and $u_{yy}$ are even in $x$ and $u_{xy}$ is odd.
        For our system, the chiral coefficients are zero and the force law commutes with reflections.
        For a configuration invariant under that mirror $\sigma_v$ we have $\mathbf{F} = \sigma_v\mathbf{F}$ and glide and torque vanish due to symmetry,
        \begin{equation}
            \label{eq:climb_null}
            \mathbf{F} \cdot \hat{\mathbf{b}} = 0,
            \qquad \tau = 0,
            \qquad \mathbf{F} = c\,\alpha_3 \, (\hat{\mathbf{z}} \times \mathbf{b}).
        \end{equation}

        These results are recovered in the simulated rafts of main text Fig.~6.
        At $\alpha_3 = 0$ the net force and torque vanish to machine precision, with $\vert \mathbf{F} \vert \le 1.4\times10^{-16}$ and $\vert \tau \vert \le 1.2 \times 10^{-15}$ in lattice units.
        $\vert \mathbf{F} \vert$ is exactly linear in $\alpha_3$, constant to $5\times10^{-12}$ when $\alpha_3$ is scaled over five decades.
        The climb force of a single dislocation takes the same value to within $1\%$ for all six Burgers orientations of the triangular lattice, with the direction of $\mathbf{F}$ within $0.9^\circ$ of the climb axis, and it reverses under $\mathbf{b} \to -\mathbf{b}$.

        The nulls of Eq.~\eqref{eq:climb_null} hold for configurations that are exactly mirror symmetric.
        A dislocation seeded from the continuum displacement field and then relaxed is only approximately so, since the branch cut of the seeding field breaks the mirror and superposing seeds compounds it.
        The glide-to-climb ratio of the net force runs from $0.015$ for a single dislocation to $0.039$, $0.060$ and $0.084$ for walls of two, three and four.
        In the trajectories of main text Fig.~6, the corresponding ratio of glide to climb velocity is $0.014$ for one dislocation and $0.081$ for four.
        These residuals are the systematic error of the construction.

        A chiral design, a ring of six dislocations whose Burgers vectors point radially outward, has no mirror but does have a six-fold rotation axis.
        That axis forces $\mathbf{F} = 0$ while leaving the torque free.
        This null is also reached only approximately.
        One of the six dislocations relaxes into a dislocation and an extra $5$--$7$ pair, which breaks the six-fold symmetry.
        The residual net force is $12\%$ of $\vert \tau \vert / r_0$, where $r_0 = 9\ell$ is the radius of the ring.

    \subsection{Seeding and relaxing dislocations}
    \label{sec:seeding}

        The raft is a triangular-lattice disk of radius $R$ lattice constants, centered on a lattice site.
        Each dislocation is seeded by displacing every particle by the continuum edge-dislocation field of Burgers vector $\mathbf{b}$, evaluated at Poisson ratio $1/3$.
        The angle of $\mathbf{b}$ is a multiple of $60^\circ$ so that it is a lattice vector, without which the defect is not topological.
        The seed supplies a starting configuration only, and the geometry that is used comes from the relaxation that follows.
        The walls are built in a raft of radius $R = 18$, with $n$ seeds stacked along the climb direction $3\ell$ apart.
        The ring is built in a raft of radius $R = 22$, with six seeds at radius $r_0 = 9\ell$ and angles $0^\circ, 60^\circ, \ldots, 300^\circ$.
        Each of its Burgers vectors points radially outward, and each branch cut runs outward to the edge.
        Then $\sum \mathbf{b}$ vanishes and the climb forces are all azimuthal.

        The seeded configuration is relaxed under harmonic springs of rest length $\ell$, minimizing $\sum_{\rm bonds} ( \vert \br_{ij} \vert - \ell )^2$ with the bond list rebuilt after each of five L-BFGS-B passes.
        The edge of the raft is free throughout.
        The triangulation is then frozen and the energy is minimized again with that fixed bond list.
        The nonreciprocal coefficient $\alpha_3$ is switched on only after this.
        The net forces and torques quoted in this section are evaluated on the relaxed geometry.
        The trajectories of main text Fig.~6 then follow from overdamped dynamics, $\gamma \dot{\mathbf{r}}_i = \mathbf{F}_i$, integrated by forward Euler for $2 \times 10^5$ steps at zero temperature with the triangulation frozen.
        The net force and torque of the relaxed raft predict the steady drift and rotation rates to within $5\%$.
        Both the seed and the relaxing springs correspond to a Poisson ratio of $1/3$, the value for a lattice of central springs.
        The measured raft has $\nu = 0.87$ to $0.93$.
        The continuum field cannot simply be evaluated there, since parts of it grow as $1/(1-\nu)$.
        At $\nu = 0.9$ its displacements reach two lattice spacings, and the relaxation produces a cluster of about eight dislocations instead of one.
        The simulated defects therefore carry the strain field of a $\nu = 1/3$ lattice.
        The symmetry results of Sec.~\ref{sec:climb} do not depend on this, but the magnitudes of the forces do.
        Lengths are in units of $\ell$, stiffnesses in units of $k_a$, and times in units of $\gamma / k_a$.

    \subsection{Robustness to a size-dependent \texorpdfstring{$\alpha_3$}{alpha3}}
    \label{sec:alpha3_size}

        The three-particle simulations show that $\alpha_3$ depends on separation, so the uniform $\alpha_3$ used above is an approximation.
        To test whether the near-cancellation among triangles survives a non-uniform coefficient, we give $\alpha_3$ a one-parameter linear profile in the mean side length of each parent triangle.
        The line passes through the measured value at $\ell$ and crosses zero at a separation $s_0$, and each triangle takes the value at its own mean side.
        Setting $s_0 \to \infty$ recovers the uniform model.
        This family is deliberately crude, since it answers whether a sign change anywhere inside the raft's triangle-size range flips the total without depending on the detailed shape of the measured profile.

        The triangles span a narrow range of sizes.
        Taking the lattice constant to be the raft-average spacing $\ell = 54.01$~$\mu$m, the mean side for the single-dislocation design is $54.0$~$\mu$m.
        The first and ninety-ninth percentiles are $53.2$ and $54.8$~$\mu$m, and the full range is $50.1$ to $56.9$~$\mu$m.
        We place the zero crossing at $45$, $50$, $55$, $58$, $60$, $63$, $66$ and $70$~$\mu$m.
        No profile tried flips the sign of the climb force.
        Placing the zero crossing at $63$~$\mu$m, where the three-particle data put it, changes the total climb force by $5\%$ for the single dislocation and by $3\%$ for a wall of four.
        The largest excursion over the whole family is a factor of $1.42$, reached when the zero crossing is placed at $55$~$\mu$m.
        That value lies just above the bulk of the triangle distribution, and anchoring the line at the measured value then forces a very steep slope across the raft, so the factor describes the one-parameter family rather than any physical prediction.

    \subsection{Robustness to a compressive pre-stress}
    \label{sec:prestress}

        The rafts of main text Fig.~6 have $a_0 = 0$ and a free edge, so their bonds carry no stress beyond the strain field of the defects.
        In the experiment the bonds are compressed, which the minimal model absorbs into $a_0 < 0$.
        To test whether the designed motion depends on this, we compress each raft with a uniform pressure applied normal to its own free edge.
        The pressure, $P = 0.28\,k_a$, acts on the boundary edges of the triangulation and moves with the raft, so it exerts no net force or torque.
        It compresses the lattice uniformly to a mean spacing $\ell' = 0.926\,\ell$, with the same spacing at the center and at the edge of the raft.

        Because the force law is linear in the bond deviations, a raft held at spacing $\ell'$ with $a_0 = 0$ is the same as a raft with reference spacing $\ell'$ and pre-stress $a_0' = k_a(\ell' - \ell) \approx -0.074\,k_a \ell$, or $a_0'/(k_a \ell') \approx -0.080$.
        The coefficient $\alpha_3$ is unchanged, since it multiplies a difference of bond deviations, which does not depend on the reference spacing.
        The compressed rafts therefore carry a compressive pre-stress of about $30\%$ of the measured value $a_0/(\ell k_a) = -0.277$.

\begin{figure}[t]
            \centering
            \includegraphics{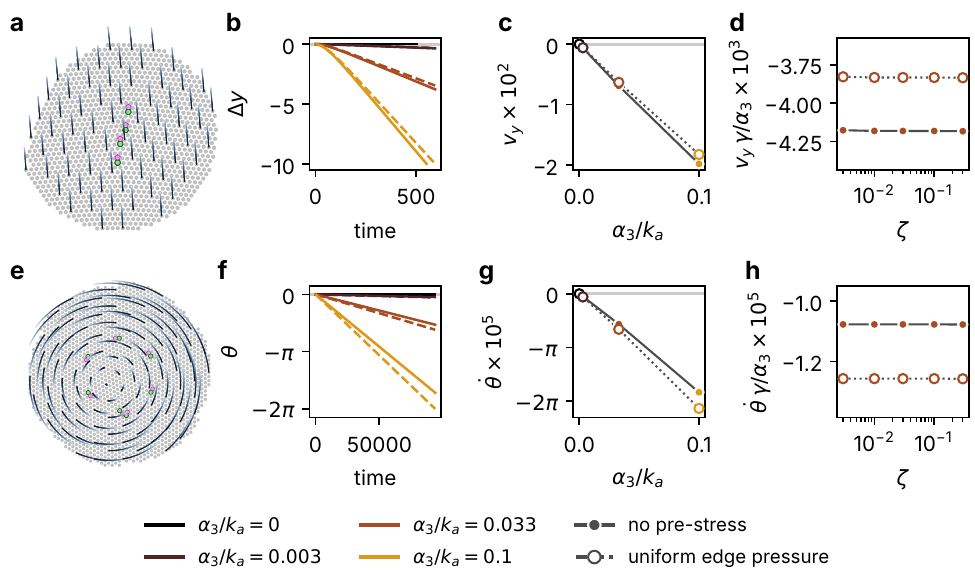}
            \caption{\textbf{Engineered activity survives a uniform pre-stress.}
            Same rafts and dynamics as main text Fig.~6: underdamped with damping ratio $\zeta = 0.01$, $a_0 = 0$, and every coefficient zero except $k_a$ and $\alpha_3$.
            Lattice units $\ell = k_a = m = 1$, with time in units of $\sqrt{m/k_a}$ and angles in radians.
            Each raft is compared without pre-stress (solid lines, filled circles) and under a uniform pressure $P = 0.28\,k_a$ applied normal to its own free edge (dashed lines, open circles), which compresses the lattice uniformly by $7.5\%$ (Sec.~\ref{sec:prestress}).
            Each raft is first relaxed under the pressure at $\alpha_3 = 0$, and $\alpha_3$ is then switched on at $t = 0$.
            Colors give $\alpha_3/k_a = 0$, $0.003$, $0.033$ and $0.1$.
            \textbf{a.} The tilt wall of four dislocations under pressure.
            5-fold and 7-fold defects are shown with magenta pentagons and green heptagons, respectively.
            Lines show the trajectories of a subset of particles while the raft travels $4\ell$ at $\alpha_3/k_a = 0.1$ (elapsed time $t \approx 270$), progressing from light to dark.
            \textbf{b.} Climb displacement $\Delta y$, in units of $\ell$, against time.
            \textbf{c.} Climb velocity $v_y$ against $\alpha_3/k_a$.
            \textbf{d.} Climb velocity rescaled by the damping coefficient, $v_y \gamma / \alpha_3$, against $\zeta$ at $\alpha_3/k_a = 0.033$.
            \textbf{e.} The ring of six dislocations under pressure, with particle trajectories over $30^\circ$ of rotation at $\alpha_3/k_a = 0.1$ (elapsed time $t \approx 7.9 \times 10^3$).
            \textbf{f.} Orientation $\theta$ against time.
            \textbf{g.} Angular velocity $\dot\theta$ against $\alpha_3/k_a$.
            \textbf{h.} Rescaled angular velocity $\dot\theta \gamma / \alpha_3$ against $\zeta$ at $\alpha_3/k_a = 0.033$.
            }
            \label{fig_si:prestress}
        \end{figure}

        Each raft is relaxed under the pressure at $\alpha_3 = 0$, after which $\alpha_3$ is switched on.
        The dynamics are otherwise identical to main text Fig.~6.
        We simulate the wall of four dislocations and the ring of six at $\alpha_3/k_a = 0$, $0.003$, $0.033$ and $0.1$ (Fig.~\ref{fig_si:prestress}).

        The pressure does not change the direction of motion of either design.
        It rescales the rates by a factor that is the same for every $\alpha_3$ and every damping ratio from $\zeta = 0.003$ to $0.3$.
        The wall climbs $8\%$ more slowly, because its net force is $9\%$ smaller.
        The ring rotates $16$--$17\%$ faster.
        Its torque is nearly unchanged, while compression reduces its rotational drag $\gamma \sum_i r_i^2$ by a factor $(\ell'/\ell)^2 = 0.86$.
        At $\alpha_3 = 0$ the compressed rafts do not move, with $\vert \mathbf{F} \vert$ and $\vert \tau \vert$ below $10^{-14}$ in lattice units.

\section{Inferring the dynamical matrix from measured trajectories}
\label{sec:inference}
    We infer the dynamical matrix of the raft from the tracked particle motion alone.
    The method assumes linear dynamics but does not assume reciprocity.
    The input is the displacement $\mathbf{u}(t)$ of all $N = 466$ particles over $62{,}999$ frames at $\delta t = 1/6000$~s.
    We first project out rigid translation and rotation of the whole raft.
    We then keep the $n_\mathrm{modes} = 40$ eigenvectors of the equal-time covariance $\langle \mathbf{u}\mathbf{u}^\mathsf{T}\rangle$ with the largest variance.
    We call these the softest modes, since in a thermal harmonic solid the softest modes fluctuate the most.
    Projecting onto them gives a 40-component time series of mode amplitudes $\mathbf{P}(t)$.
    The velocities $\dot{\mathbf{P}}$ come from a third-order Savitzky--Golay derivative over nine frames.
    We stack the two into $\mathbf{z} = [\mathbf{P}, \dot{\mathbf{P}}]$ and model $\dot{\mathbf{z}} = -\mathcal{J}\mathbf{z} + \boldsymbol{\xi}$ with white noise $\boldsymbol{\xi}$.
    Over one frame this gives
    \begin{equation}
    \label{eq:propagator}
        e^{-\mathcal{J}\delta t} = \langle \mathbf{z}(t+\delta t)\,\mathbf{z}^\mathsf{T}(t)\rangle \, \langle \mathbf{z}\mathbf{z}^\mathsf{T}\rangle^{-1},
        \qquad
        \mathcal{J} = \begin{pmatrix} 0 & -I \\ D & \boldsymbol{\gamma} \end{pmatrix}.
    \end{equation}
    We take the matrix logarithm of the right-hand side and read $D$ from the lower-left block.
    Nothing in this construction forces $D$ to be symmetric.
    Its antisymmetric part $D^-$ is therefore free to take any value the data support.

    Tracking error adds white noise to every measured position.
    It appears as a flat floor above 900~Hz in the power spectrum of every mode.
    The floor corresponds to a noise amplitude of $0.18~\mu$m, the same in all 40 modes to within $\pm 5\%$.
    That is $0.6\%$ of the fluctuation of the softest mode and $8\%$ of the fluctuation of the fortieth.
    The Savitzky--Golay derivative carries this noise into $\dot{\mathbf{P}}$, where it could in principle mimic an antisymmetric $D$.
    We test this on simulated reciprocal rafts with 40 modes between 5 and 228~Hz and damping rates between 25 and 270~s$^{-1}$.
    The simulations have the length and frame rate of the experiment and carry isotropic white noise at the measured level.
    Each is passed through the full pipeline, including the choice of mode basis.
    They return $\Vert D^- \Vert / \Vert D^+ \Vert = 0.012$ to $0.017$, against $0.204$ for the raft.
    This holds when the damping couples the modes and when the noise is raised to $25\%$ of a mode's fluctuation.
    The result relies on the noise being isotropic.
    Noise of strongly unequal size in directions that mix the modes can produce a ratio comparable to the measured one.
    The measured floor rules this out.

    Under detailed balance the lagged correlations of the mode amplitudes obey $C_{ij}(\tau) = C_{ji}(\tau)$.
    Equivalently, their cross-spectral matrix $\tilde{C}(f)$ is real.
    Its imaginary part therefore carries all of the time irreversibility.
    We build a reciprocal null by keeping only $\mathrm{Re}\,\tilde{C}(f)$.
    We estimate $\tilde{C}(f)$ by Welch averaging over 58 Hann-windowed segments of 4096 frames with $75\%$ overlap.
    We then draw Gaussian time series whose cross-spectral matrix is $\mathrm{Re}\,\tilde{C}(f)$, with the length of the experiment.
    Each surrogate keeps the power spectrum of every mode, including its spectral lines and tracking-noise floor.
    It also keeps every reciprocal coupling between modes.
    Passing 40 surrogates through the same pipeline gives $\Vert D^- \Vert / \Vert D^+ \Vert = 0.121 \pm 0.001$, against $0.204$ for the raft.
    The measured value lies 81 standard deviations above the null.

    We test the null in two ways.
    Surrogates drawn from the full complex $\tilde{C}(f)$ keep the irreversible part and should reproduce the measured value.
    They give $0.204 \pm 0.001$, against $0.204$ for the raft.
    This control agrees with the measurement to within $2\%$ at every mode count in Table~\ref{tab:inference_M}.
    We also apply the full procedure to simulated rafts whose dynamical matrix is known.
    Each has 40 coupled modes between 5 and 260~Hz, a damping rate of $189$~s$^{-1}$, and an antisymmetric part of chosen strength.
    The simulations have the length and frame rate of the experiment.
    A reciprocal raft and a raft with true $\Vert D^- \Vert / \Vert D^+ \Vert = 0.026$ both fall inside their nulls.
    Rafts with true ratios of $0.064$ and $0.128$ are detected.
    In both detected cases the pipeline returns $0.59$ of the true ratio.
    We have not shown that this factor carries over to the raft, so we do not correct for it.
    Nor can the null be subtracted.
    Subtracting its mean in quadrature returns $0.049$ for a true ratio of $0.128$.
    We therefore report the measured value alongside its null.

    The surrogates carry $2.4\%$ of the power of the data.
    This offset comes from how they are drawn.
    $\tilde{C}(f)$ is estimated on Hann-tapered windows of 4096 frames, and full-length series are drawn from it without rescaling.
    That reduces the power by $\sum_n w_n^2 / T = 0.0244$, where $w_n$ is the taper and $T = 62{,}999$ is the number of frames.
    The measured power ratio is $0.0237$ averaged over modes, and between $0.021$ and $0.026$ for individual modes.
    The shape of each mode's spectrum is reproduced, with a mean correlation of $0.997$ between the logarithms of the two spectra.
    The offset does not enter the inferred $D$.
    Equation~\eqref{eq:propagator} divides one covariance of $\mathbf{z}$ by another, so a common rescaling of all amplitudes cancels.
    Restoring the surrogates to full power leaves $\Vert D^- \Vert / \Vert D^+ \Vert$ unchanged to six digits.
    The spread between modes does not cancel exactly.
    Imposing it on the measured trajectories moves $\Vert D^- \Vert / \Vert D^+ \Vert$ from $0.2036$ to $0.2046$, a change of $0.5\%$.

    The result depends on how many modes are kept (Table~\ref{tab:inference_M}).
    The raft exceeds its null at every mode count.
    The excess is largest at $n_\mathrm{modes} = 10$ and $n_\mathrm{modes} = 40$ and smallest at $n_\mathrm{modes} = 160$, where the raft gives $0.201$ against a null of $0.186$.
    It does not fall monotonically with $n_\mathrm{modes}$, and we have no explanation for this dependence.
    We use $n_\mathrm{modes} = 40$ throughout.

    \begin{table}[h]
        \centering
        \caption{Inferred $\Vert D^- \Vert / \Vert D^+ \Vert$ against the number of modes $n_\mathrm{modes}$.
        The null is the mean and standard deviation over 40 reciprocal surrogates.
        The control uses surrogates drawn from the full complex $\tilde{C}(f)$, averaged over 6 realizations.}
        \label{tab:inference_M}
        \begin{tabular}{r c c c c}
            \hline\hline
            $n_\mathrm{modes}$ & raft & null & raft / null & control \\
            \hline
            10  & 0.155 & $0.077 \pm 0.002$ & 2.01 & 0.157 \\
            20  & 0.177 & $0.120 \pm 0.001$ & 1.48 & 0.180 \\
            40  & 0.204 & $0.121 \pm 0.001$ & 1.68 & 0.204 \\
            80  & 0.192 & $0.157 \pm 0.001$ & 1.22 & 0.192 \\
            160 & 0.201 & $0.186 \pm 0.001$ & 1.08 & 0.205 \\
            \hline\hline
        \end{tabular}
    \end{table}

\section{Position space circulation is forbidden in underdamped systems in equilibrium}
    Consider the underdamped Langevin equation with white noise
    \begin{equation}
        \partial_{tt} \br = - \gamma \partial_t \br + \mathbf{F}(\br) + \sqrt{2 T \gamma} \xi; \quad \langle \xi(t) \xi(t') \rangle = \delta(t-t'),
    \end{equation}
    where $\mathbf{F}(\br) = - \nabla_\br U$ is conservative and the mass is set to unity.
    The probability distribution obeys the Kramers--Chandrasekhar Fokker--Planck equation
    \begin{equation}
    \label{eq:kcfp}
        \left(\partial_t + \mathbf{v} \cdot \nabla_\br \right)p(\br, \mathbf{v}) = \nabla_\mathbf{v} \cdot \left[ \left( \gamma \mathbf{v} - \mathbf{F}(\br) \right) p \right] + T \gamma \, \nabla^2_\mathbf{v} p,
    \end{equation}
    with a steady-state solution $p_\mathrm{eq}(\mathbf{r}, \mathbf{v}) \propto \exp(-E/T)$, where the energy is $E(\br, \mathbf{v}) = U(\br) + v^2 / 2$.

    We can rewrite Eq.~\eqref{eq:kcfp} as $\partial_t p = - \nabla_\br \cdot \mathbf{J}_\br - \nabla_\mathbf{v} \cdot \mathbf{J}_\mathbf{v}$, where
    \begin{equation}
        \begin{pmatrix}
            \mathbf{J}_\br(\br, \mathbf{v}) \\
            \mathbf{J}_\mathbf{v}(\br, \mathbf{v})
        \end{pmatrix}
        = 
        \begin{pmatrix}
            \mathbf{v} p \\
            (\mathbf{F}(\br) - \gamma \mathbf{v}) p - \gamma T \, \nabla_\mathbf{v} p
        \end{pmatrix}
        \overset{\mathrm{eq}}{=}
        \begin{pmatrix}
            \mathbf{v} \\
            \mathbf{F}
        \end{pmatrix} p_\mathrm{eq},
    \end{equation}
    where the last equation is true in equilibrium. 
    Marginalizing $\mathbf{J}_\br$ over the velocity, we find
    \begin{equation}
        \mathbf{J}_\br(\br) = \int \mathbf{J}_\br(\br, \mathbf{v}) \, d\mathbf{v} = \int \mathbf{v} \, p_\mathrm{eq}(\br, \mathbf{v}) \, d\mathbf{v} = 0,
    \end{equation}
    where the last equality follows from $p_\mathrm{eq}$ being even in $\mathbf{v}$.
    Thus, we see that not only is $\nabla_\br \times \mathbf{J}_\br = 0$, but also $\mathbf{J}_\br = 0$.

\twocolumngrid

\end{document}